\documentclass[journal,final,twoside]{IEEEtran}
\usepackage{cite}

\ifCLASSINFOpdf
   \usepackage[pdftex]{graphicx}
   \graphicspath{{./figures/}{./pdf/}{./jpeg/}}
   \DeclareGraphicsExtensions{.pdf,.jpeg,.png}
\else
   \usepackage[dvips]{graphicx}
   \graphicspath{{./eps/}}
   \DeclareGraphicsExtensions{.eps}
\fi
\usepackage{algorithm}
\usepackage{algpseudocode}

\usepackage[cmex10]{amsmath}
\usepackage{amssymb}

\usepackage[caption=false,font=footnotesize]{subfig}
\usepackage{stfloats}

\usepackage{multirow}

\usepackage{footnote}

\usepackage{threeparttable}

\renewcommand{\vec}[1]{\mathbf{#1}}
\newcommand{\vecRvReal}[1]{{#1}}

\newcommand{\mtx}[1]{\mathbf{#1}}

\newtheorem{problemform}{Problem}

\begin{document}
%
\title{Restoration by Compression}
%
%
%

\author{Yehuda Dar, Michael Elad, and Alfred M. Bruckstein
\\
\thanks{The authors are with the Department of Computer Science, Technion, Israel. E-mail addresses: \{ydar,~elad,~freddy\}@cs.technion.ac.il.}
}

%
%

\markboth{}%
{~}
%



\maketitle

\begin{abstract}
In this paper we study the topic of signal restoration using complexity regularization, quantifying the compression bit-cost of the signal estimate. 
While complexity-regularized restoration is an established concept, solid practical methods were suggested only for the Gaussian denoising task, leaving more complicated restoration problems without a generally constructive approach. 
Here we present practical methods for complexity-regularized restoration of signals, accommodating deteriorations caused by a known linear degradation operator of an arbitrary form.
Our iterative procedure, obtained using the alternating direction method of multipliers (ADMM) approach, addresses the restoration task as a sequence of simpler problems involving $ \ell _2$-regularized estimations and rate-distortion optimizations (considering the squared-error criterion). Further, we replace the rate-distortion optimizations with an arbitrary standardized compression technique and thereby restore the signal by leveraging underlying models designed for compression.
Additionally, we propose a shift-invariant complexity regularizer, measuring the bit-cost of all the shifted forms of the estimate, extending our method to use averaging of decompressed outputs gathered from compression of shifted signals.
On the theoretical side, we present an analysis of complexity-regularized restoration of a cyclo-stationary Gaussian signal from deterioration by a linear shift-invariant operator and an additive white Gaussian noise. The theory shows that optimal complexity-regularized restoration relies on an elementary restoration filter and compression spreading reconstruction quality unevenly based on the energy distribution of the degradation filter. Nicely, these ideas are realized also in the proposed practical methods.
Finally, we present experiments showing good results for image deblurring and inpainting using the JPEG2000 and HEVC compression standards. 
\end{abstract}


~\\~

%
\IEEEpeerreviewmaketitle

\section{Introduction}
\IEEEPARstart{S}{ignal} restoration methods are often posed as inverse problems using regularization terms. While many solutions can explain a given degraded signal, using regularization will provide signal estimates based on prior assumptions on signals.
One interesting regularization type measures the complexity of the candidate solution in terms of its compression bit-cost. Indeed, encoders (that yield the bit cost) rely on signal models and allocate shorter representations to more likely signal instances. 
This approach of complexity-regularized restoration is an attractive meeting point of signal restoration and compression, two fundamental signal-processing problems.

Numerous works \cite{saito1994simultaneous,natarajan1995filtering,chang1997image,mihcak1999low,rissanen2000mdl,chang2000adaptive,liu2001complexity} considered the task of denoising a signal corrupted by an additive white Gaussian noise using complexity regularization. In \cite{natarajan1995filtering,liu2001complexity}, this idea is translated to practically estimating the clean signal by employing a standard lossy compression of its noisy version. 
However, more complex restoration problems (e.g., deblurring, super resolution, inpainting), involving non-trivial degradation operators, do not lend themselves to a straightforward treatment by compression techniques designed for the squared-error distortion measure.
Moulin and Liu \cite{moulin2000statistical} studied the complexity regularization idea for general restoration problems, presenting a thorough theoretical treatment together with a limited practical demonstration of Poisson denoising based on a suitably designed compression method.
Indeed, a general method for complexity-regularized restoration remained as an open question for a long while until our recent preliminary publication \cite{dar2016image}, where we presented a generic and practical approach flexible in both the degradation model addressed and the compression technique utilized.

Our strategy for complexity-regularized signal restoration relies on the alternating direction method of multipliers (ADMM) approach \cite{boyd2011distributed}, decomposing the difficult optimization problem into a sequence of easier tasks including $ \ell_2 $-regularized inverse problems and standard rate-distortion optimizations (with respect to a squared-error distortion metric). 
A main part of our methodology is to replace the rate-distortion optimization with standardized compression techniques enabling an indirect utilization of signal models used for efficient compression designs.
Moreover, our method relates to various contemporary concepts in signal and image processing. The recent frameworks of Plug-and-Play Priors \cite{venkatakrishnan2013plug,sreehari2016plug} and Regularization-by-Denoising \cite{romano2017little} suggest leveraging a Gaussian denoiser for more complicated restoration tasks, achieving impressive results (see, e.g., \cite{venkatakrishnan2013plug,sreehari2016plug,romano2017little,dar2016postprocessing,dar2016reducing,rond2016poisson}). Essentially, our approach is the compression-based counterpart for denoising-based restoration concepts from \cite{venkatakrishnan2013plug,sreehari2016plug,romano2017little}. 

Commonly, compression methods process the given signal based on its decomposition into non-overlapping blocks, yielding block-level rate-distortion optimizations based on block bit-costs. The corresponding complexity measure sums the bit-costs of all the non-overlapping blocks, however, note that this evaluation is shift sensitive. This fact motivates us to propose a shift-invariant complexity regularizer by quantifying the bit-costs of all the overlapping blocks of the signal estimate. 
This improved regularizer calls for our restoration procedure to use averaging of decompressed signals obtained from compressions of shifted signals.
Our shift-invariant approach conforms with the Expected Patch Log-Likelihood (EPLL) idea \cite{zoran2011learning}, where a full-signal regularizer is formed based on a block-level prior in a way leading to averaging MAP estimates of shifted signal versions.
Our extended method also recalls the cycle spinning concept, presented in \cite{coifman1995translation} for wavelet-based denoising. 
Additional resemblance is to the compression postprocessing techniques in \cite{nosratinia2001enhancement,nosratinia2003postprocessing} enhancing a given decompressed image by averaging supplementary compression-decompression results of shifted versions of the given image, thus, our method generalizes this approach to any restoration problem with an appropriate consideration of the degradation operator.
Very recent works \cite{beygi2017compressed,beygi2017efficient} suggested the use of compression techniques for compressive sensing of signals and images, but our approach examines other perspectives and settings referring to restoration problems as will be explained below.

In this paper we extend our previous conference publication \cite{dar2016image} with improved algorithms and new theoretical and experimental results. 
In \cite{dar2016image} we implemented our concepts in procedures relying on the half quadratic splitting optimization technique, in contrast, here we present improved algorithms designed based on the ADMM approach. The new ADMM-based methods introduce the following benefits (with respect to using half quadratic splitting as in \cite{dar2016image}): significant gains in the restoration quality, reduction in the required amount of iterations, and an easier parameter setting.
In addition, in this paper we provide an extensive experimental section. While in \cite{dar2016image} we experimentally examined only the inpainting problem, in this paper we present new results demonstrating the practical complexity-regularized restoration approach for image deblurring. While deblurring is a challenging restoration task, we present compelling results obtained using the JPEG2000 method and the image compression profile of the HEVC standard \cite{RefWorks:112}. An objective comparison to other deblurring techniques showed that the proposed HEVC-based implementation provides good deblurring results. 
Moreover, we also extend our evaluation given in \cite{dar2016image} for image inpainting, where here we use the JPEG2000 and HEVC compression standards in our ADMM-based approach to restore images from a severe degradation of 80\% missing pixels.
Interestingly, our compression-based image inpainting approach can be perceived as the dual concept of inpainting-based compression of images and videos suggested in, e.g., \cite{galic2008image,schmaltz2009beating,andris2016proof} and discussed also in \cite{adam2017denoising}.

Another prominent contribution of this paper is the new theoretical study of the problem of complexity-regularized restoration, considering the estimation of a cyclo-stationary Gaussian signal from a degradation procedure consisting of a linear shift-invariant operator and additive white Gaussian noise. 
We gradually establish few equivalent optimization forms, emphasizing two main concepts for complexity-regularized restoration: the degraded signal should go through a simple inverse filtering procedure, and then should be compressed so that the decompression components will have a varying quality distribution determined by the degradation-filter energy-distribution. We explain how these ideas materialize in the practical approach we propose, thus, establishing a theoretical reasoning for the feasible complexity-regularized restoration.

This paper is organized as follows. In section \ref{sec:Complexity-Regularized Restoration} we overview the settings of the complexity-regularized restoration problem.
In section \ref{sec:Proposed Methods} we present the proposed practical methods for complexity-regularized restoration.
In section \ref{sec:Rate-Distortion Theoretic Analysis for the Gaussian Case} we theoretically analyze particular problem settings where the signal is a cyclo-stationary Gaussian process.
In section \ref{sec:Experimental Results} we provide experimental results for image deblurring and inpainting.
Section \ref{sec:Conclusion} concludes this paper.

\section{Complexity-Regularized Restoration: Problem Settings}
\label{sec:Complexity-Regularized Restoration}

\subsection{Regularized Restoration of Signals}
\label{subsec:Regularized-Restoration Optimization}
In this paper we address the task of restoring a signal $ \vec{x}_0 \in\nolinebreak \mathbb{R}^N $ from a degraded version, $ \vec{y} \in \mathbb{R}^M $, obeying the prevalent deterioration model: 
\begin{IEEEeqnarray}{rCl}
	\label{eq:corruption model}
	{\vec{y}} = \mtx{H} \vec{x}_0 + \vec{n}
\end{IEEEeqnarray}
where $ \mtx{H} $ is a $M\times N$ matrix being a linear degradation operator (e.g., blur, pixel omission, decimation) and $\vec{n}\in \mathbb{R}^M$ is a white Gaussian noise vector having zero mean and variance $\sigma _n ^2$.

Maximum A-Posteriori (MAP) estimation is a widely-known statistical approach forming the restored signal, $\hat{ \vec{x}}$, via 
\begin{IEEEeqnarray}{rCl}
	\label{eq:MAP estimation}
	\hat{ \vec{x}} = \mathop {{\text{argmax}}}\limits_{\vec{x}} p\left( {{\vec{x}}|{\vec{y}}} \right)
\end{IEEEeqnarray}
where $  p\left( {{\vec{x}}|{\vec{y}}} \right) $ is the posterior probability. For the above defined degradation model (\ref{eq:corruption model}), incorporating additive white Gaussian noise, the MAP estimate reduces to the form of 
\begin{IEEEeqnarray}{rCl}
	\label{eq:log MAP estimate - AWGN model}
	\hat{ \vec{x}} = \mathop {{\text{argmin}}}\limits_{\vec{x}} \frac{1}{2\sigma ^2_n} \left\| {  \mtx{H} \vec{x} - \vec{y} } \right\|_2^2 - \log p \left( {{\vec{x}}} \right)
\end{IEEEeqnarray}
where $p(\vec{x})$ is the prior probability that, here, evaluates the probability of the candidate solution.

Another prevalent restoration approach, embodied in many contemporary techniques, forms the estimate via the optimization 
\begin{IEEEeqnarray}{rCl}
	\label{eq:restoration using a general prior}
	\hat{ \vec{x}} = \mathop {{\text{argmin}}}\limits_{\vec{x}} \left\| {  \mtx{H} \vec{x} - \vec{y} } \right\|_2^2 + \mu s(\vec{x})
\end{IEEEeqnarray}
where $s(\vec{x})$ is a general regularization function returning a lower value for a more likely candidate solution, and $\mu\ge 0$ is a parameter weighting the regularization effect. 
This strategy for restoration based on arbitrary regularizers can be interpreted as a generalization of the MAP approach in (\ref{eq:log MAP estimate - AWGN model}). Specifically, comparing the formulations (\ref{eq:restoration using a general prior})  and (\ref{eq:log MAP estimate - AWGN model}) exhibits the regularization function $s(\vec{x})$ and the parameter $ \mu $ as extensions of $( - \log p \left( {{\vec{x}}} \right)) $ and the factor $2\sigma ^2_n$, respectively.

Among the various regularization functions that can be associated with the general restoration approach in (\ref{eq:restoration using a general prior}), we explore here the class of complexity regularizers measuring the required number of bits for the compressed representation of the candidate solution. The practical methods presented in this section focus on utilizing existing (independent) compression techniques, implicitly employing their underlying signal models for the restoration task.

\subsection{Operational Rate-Distortion Optimization}
\label{subsec:Operational Rate-Distortion Optimization}
The practical complexity-regularized restoration methods in this section are developed with respect to a compression technique obeying the following conceptual design. The signal is segmented to equally-sized non-overlapping blocks (each is consisted of $ N_b $ samples) that are independently compressed. 
The block compression procedure is modeled as a general variable-rate vector quantizer relying on the following mappings. The compression is done by the mapping $ Q : \mathbb{R}^{N_b}  \rightarrow \mathcal{W} $ from the $ N_b $-dimensional signal-block domain to a discrete set $ \mathcal{W} $ of binary compressed representations (that may have different lengths). The decompression procedure is associated with the mapping $ F :  \mathcal{W} \rightarrow \mathcal{C} $, where $ \mathcal{C} \subset \mathbb{R}^{N_b} $ is a finite discrete set (a codebook) of block reconstruction candidates. For example, consider the block $\vec{x}_{block} \in \mathbb{R}^{N_b}$ that its binary compressed representation in $ \mathcal{W} $ is given via $ b = Q\left( \vec{x}_{block} \right) $ and the corresponding reconstructed block in $ \mathcal{C} $ is $ \hat{\vec{x}}_{block} = F\left(  b \right) $.
Importantly, it is assumed that shorter codewords are coupled with block reconstructions that are, in general, more likely.

The signal $ \vec{x} $ is compressed based on its segmentation into a set of blocks $\{\vec{x}_i\}_{i\in\mathcal{B}}$ (where $ \mathcal{B}$ denotes the index set of blocks in the non-overlapping partitioning of the signal). In addition we introduce the function  $ r(\vec{z}) $ that evaluates the bit-cost (i.e., the length of the binary codeword) for the block reconstruction $ \vec{z}\in\mathcal{C} $.
Then, the operational rate-distortion optimization corresponding to the described architecture and a squared-error distortion metric is 
\begin{IEEEeqnarray}{rCl}
	\label{eq:rate-distortion optimization - blocks}
	\{\vec{\tilde x}_i\}_{i\in\mathcal{B}} = \mathop {{\text{argmin}}}\limits_{\{\vec{v}_i\}_{i\in\mathcal{B}} \in \mathcal{C}}  \mathop \sum\limits_{i\in\mathcal{B}}{ \left\| {  \vec{x}_i - \vec{v}_i } \right\|_2^2} + \lambda \sum\limits_{i\in\mathcal{B}}{r(\vec{v}_i)} ,
\end{IEEEeqnarray}
where $ \lambda \ge 0 $ is a Lagrange multiplier corresponding to some total compression bit-cost. 
Importantly, the independent representation of non-overlapping blocks allows solving (\ref{eq:rate-distortion optimization - blocks}) separately for each block \cite{shoham1988efficient,ortega1998rate}.

Our mathematical developments require the following algebraic tools for block handling. The matrix $ \mtx{P}_i $ is defined to provide the $ i^{th} $ block from the complete signal via the standard multiplication $\mtx{P}_i\vec{x} = \vec{x}_i$. Note that $ \mtx{P}_i $ can extract any block of the signal, even one that is not in the non-overlapping grid $ \mathcal{B} $. Accordingly, the matrix $ \mtx{P}_i^T $ locates a block in the $ i^{th} $ block-position in a construction of a full-sized signal and, therefore, lets to express the a complete signal as $\vec{x} = \mathop \sum\limits_{i\in\mathcal{B}} \mtx{P}_i^T\vec{x}_i$.

Now we can use the block handling operator $ \mtx{P}_i $ for expressing the block-based rate-distortion optimization in its corresponding full-signal formulation: 
\begin{IEEEeqnarray}{rCl}
	\label{eq:rate-distortion optimization - full signal}
	\tilde{\vec{x}} = \mathop {{\text{argmin}}}\limits_{\vec{v} \in \mathcal{C}_{\mathcal{B}} }  { \left\| {  \vec{x} - \vec{v} } \right\|_2^2} + \lambda r_{tot}(\vec{v}) .
\end{IEEEeqnarray}
where $ \mathcal{C}_{\mathcal{B}} $ is the full-signal codebook, being the discrete set of candidate reconstructions for the full signal, defined using the block-level codebook $ \mathcal{C} $ as 
\begin{IEEEeqnarray}{rCl}
	\label{eq:group of solutions based on nonoverlapping blocks}
	\mathcal{C}_{\mathcal{B}} = \left\lbrace {\vec{v} ~~ \Big|{~~ {\vec{v} = \mathop \sum\limits_{i\in\mathcal{B}}{\mtx{P}_i^T \vec{v}_i}, ~~  {\{\vec{v}_i\}_{i\in\mathcal{B}} \in \mathcal{C}}} } }\right\rbrace.
\end{IEEEeqnarray}
Moreover, the regularization function in (\ref{eq:rate-distortion optimization - full signal}) is the total bit cost of the reconstructed signal defined for $ \vec{v} \in \mathcal{C}_{\mathcal{B}} $ as $	r_{tot}(\vec{v}) \triangleq \nolinebreak \sum\limits_{i\in \mathcal{B}} r(\mtx{P}_i \vec{v})$.

\subsection{Complexity-Regularized Restoration: Basic Optimization Formulation}
\label{subsec:Complexity-Regularized Restoration}

While the regularized-restoration optimization in (\ref{eq:restoration using a general prior}) is over a continuous domain, the operational rate-distortion optimization in (\ref{eq:rate-distortion optimization - full signal}) is a discrete problem with solutions limited to the set $ \mathcal{C}_{\mathcal{B}} $. 
Therefore, we extend the definition of the block bit-cost evaluation function such that it is defined for any $ \vec{z} \in \mathbb{R}^{N_b} $ via 
\begin{IEEEeqnarray}{rCl}
	\label{eq:extended block bit-cost}
	\bar r (\vec{z}) = \left\{ {\begin{array}{*{20}{c}}
			{r\left( \vec{z} \right)}&{,\vec{z} \in \mathcal{C}} \\ 
			\infty &{,\vec{z} \notin \mathcal{C}} 
		\end{array}} \right. ,
\end{IEEEeqnarray}
and the corresponding extension of the total bit-cost $\bar{r}_{tot}(\vec{x})\triangleq\nolinebreak \sum\limits_{i\in \mathcal{B}} \bar r(\mtx{P}_i \vec{x})$ is defined for any  $ \vec{x} \in \mathbb{R}^{N} $. 

Now we define the complexity regularization function as 
\begin{IEEEeqnarray}{rCl}
	\label{eq:complexity regularization function}
s(\vec{x})\nolinebreak=\nolinebreak\bar{r}_{tot}(\vec{x})
\end{IEEEeqnarray}
and the corresponding restoration optimization is 
\begin{IEEEeqnarray}{rCl}
	\label{eq:complexity-regularized restoration - full signal}
	\hat{ \vec{x}} = \mathop {{\text{argmin}}}\limits_{\vec{x}} \left\| {  \mtx{H} \vec{x} - \vec{y} } \right\|_2^2 + \mu \bar{r}_{tot}(\vec{x}).
\end{IEEEeqnarray}
Due to the definition of the extended bit-cost evaluation function, $\bar{r}_{tot}(\vec{x})$, the solution candidates of (\ref{eq:complexity-regularized restoration - full signal}) are limited to the discrete set $ \mathcal{C}_{\mathcal{B}}  $ as defined in (\ref{eq:group of solutions based on nonoverlapping blocks}).

Examining the complexity-regularized restoration in (\ref{eq:complexity-regularized restoration - full signal}) for the Gaussian denoising task, where $ \mtx{H} = \mtx{I} $, shows that the optimization reduces to the regular rate-distortion optimization in (\ref{eq:rate-distortion optimization - full signal}), namely, the compression of the noisy signal $ \vec{y} $. However, for more complicated restoration problems, where $ \mtx{H} $ has an arbitrary structure, the optimization in (\ref{eq:complexity-regularized restoration - full signal}) is not easy to solve and, in particular, it does not correspond to standard compression designs that are optimized for the regular squared-error distortion metric.

\section{Proposed Methods}
\label{sec:Proposed Methods}

In this section we present three restoration methods leveraging a given compression technique. The proposed algorithms result from two different definitions for the complexity regularization function. While the first approach regularizes the total bit-cost of the non-overlapping blocks of the restored signal, the other two refer to the total bit-cost of all the overlapping blocks of the estimate.

\subsection{Regularize Total Complexity of Non-Overlapping Blocks}
\label{subsec:Total Complexity of Non-Overlapping Blocks}

Here we establish a practical method addressing the optimization problem in (\ref{eq:complexity-regularized restoration - full signal}) based on the alternating direction method of multipliers (ADMM) approach \cite{boyd2011distributed} (for additional uses see, e.g., \cite{venkatakrishnan2013plug,sreehari2016plug,dar2016postprocessing,rond2016poisson,afonso2010fast}).
The optimization (\ref{eq:complexity-regularized restoration - full signal}) can be expressed also as 
\begin{IEEEeqnarray}{rCl}
	\label{eq:complexity-regularized restoration - expressing blocks}
	\hat{ \vec{x}} = \mathop {{\text{argmin}}}\limits_{\vec{x}} \left\| {  \mtx{H} \vec{x} - \vec{y} } \right\|_2^2 + \mu  \sum\limits_{i\in \mathcal{B}} \bar r(\mtx{P}_i\vec{x}),
\end{IEEEeqnarray}
where the degradation matrix $ \mtx{H} $, having a general structure, renders a block-based treatment infeasible.

Addressing this structural difficulty using the ADMM strategy \cite{boyd2011distributed} begins with introducing the auxiliary variables $ \left\lbrace \vec{z}_i \right\rbrace_{i\in\mathcal{B}} $, where $ \vec{z}_i $ is coupled with the $ i^{th} $ non-overlapping block. Specifically, we reformulate the problem (\ref{eq:complexity-regularized restoration - expressing blocks}) into
\begin{IEEEeqnarray}{rCl}
	\label{eq:complexity-regularized restoration - with splitting - constrained form}
	&&\left( \hat{ \vec{x}}, \left\lbrace \hat{\vec{z}}_i \right\rbrace_{i\in\mathcal{B}} \right) = \mathop {{\text{argmin}}}\limits_{\vec{x},\left\lbrace \vec{z}_i \right\rbrace_{i\in\mathcal{B}}} \left\| {  \mtx{H} \vec{x} - \vec{y} } \right\|_2^2 + 
	\mu  \mathop\sum\limits_{i\in \mathcal{B}} {\bar{r}(\vec{z}_i)} \nonumber\\ 
	&& \text{s.t. }~~~~~ \vec{z}_i = \mtx{P}_i \vec{x} ~\text{,~~for~} i\in\mathcal{B} .
\end{IEEEeqnarray}
Then, reformulating the constrained optimization (\ref{eq:complexity-regularized restoration - with splitting - constrained form}) using the augmented Lagrangian (in its scaled form) and the method of multipliers (see \cite[Ch. 2]{boyd2011distributed}) leads to the following iterative procedure
\begin{IEEEeqnarray}{rCl}
	\label{eq:complexity-regularized restoration - with splitting}
	\left( \hat{ \vec{x}}^{(t)}, \left\lbrace \hat{\vec{z}}_i^{(t)} \right\rbrace_{i\in\mathcal{B}} \right) &=& \mathop {{\text{argmin}}}\limits_{\vec{x},\left\lbrace \vec{z}_i \right\rbrace_{i\in\mathcal{B}}} \left\| {  \mtx{H} \vec{x} - \vec{y} } \right\|_2^2 + \mu  \sum\limits_{i\in \mathcal{B}} \bar r(\vec{z}_i) ~~~~~~\nonumber\\ 
	&& \qquad + \frac{\beta}{2}\sum\limits_{i\in \mathcal{B}} {\left\| {  \mtx{P}_i \vec{x} - \vec{z}_i + \vec{u}_i^{(t) } } \right\|_2^2}  
	\\
	\vec{u}_i^{(t+1) } & = & \vec{u}_i^{(t) } + \left( \mtx{P}_i \hat{\vec{x}}^{(t)} - \hat{\vec{z}}_i^{(t)} \right)	~~ i\in\mathcal{B} , 
\end{IEEEeqnarray}
where $ t $ is the iteration number, $ \beta $ is a parameter originating in the augmented Lagrangian, and $ \vec{u}_i^{(t) } \in \mathbb{R}^{N_b} $ is the scaled dual variable corresponding to the $ i^{th} $ block (where $ i\in \mathcal{B} $).

Each of the optimization variables in (\ref{eq:complexity-regularized restoration - with splitting}) participates only in part of the terms of the cost function and, therefore, employing one iteration of alternating minimization (see \cite[Ch. 2]{boyd2011distributed}) leads to the ADMM form of the problem, where the included optimizations are relatively simple.
Accordingly, the $ t^{th} $ iteration of the proposed iterative solution is
\begin{IEEEeqnarray}{rCl}
	\label{eq:complexity-regularized restoration - iterative solution - inversion}
	&&  \hat{ \vec{x}}^{(t)} = \mathop {\text{argmin}}\limits_{\vec{x}} \left\| {  \mtx{H} \vec{x} - \vec{y} } \right\|_2^2 + \frac{\beta}{2} \sum\limits_{i\in \mathcal{B}} {\left\| {  \mtx{P}_i \vec{x} - {\tilde{\vec{z}}}_{i}^{(t)}} \right\|_2^2}~~~~   \\ 
	\label{eq:complexity-regularized restoration - iterative solution - compression}
	&&  \hat{\vec{z}}_i^{(t)} = \mathop {\text{argmin}}\limits_{\vec{z}_i} \frac{\beta}{2}   {\left\| {  {\tilde{\vec{x}}}_{i}^{(t)} - \vec{z}_i } \right\|_2^2}   + \mu   \bar r(\vec{z}_i),   ~~ i\in\mathcal{B} \\
	\label{eq:complexity-regularized restoration - iterative solution - beta update}
	&& 	\vec{u}_i^{(t+1)} = \vec{u}_i^{(t)} + \left( \mtx{P}_i \hat{\vec{x}}^{(t)} - \hat{\vec{z}}_i^{(t)} \right)	,   ~~ i\in\mathcal{B} , 
\end{IEEEeqnarray}
where $ {\tilde{\vec{z}}}_{i}^{(t)} \triangleq {\hat{\vec{z}}}_{i}^{(t-1)} - \vec{u}_i^{(t) } $ and $ {\tilde{\vec{x}}}_{i}^{(t)} \triangleq \mtx{P}_i \hat{\vec{x}}^{(t)} + \vec{u}_i^{(t)} $  for $  i\in\mathcal{B} $.

The analytic solution of the first stage optimization in (\ref{eq:complexity-regularized restoration - iterative solution - inversion}) is 
\begin{IEEEeqnarray}{rCl}
	\label{eq:complexity-regularized restoration - iterative solution - inversion - analytic form}
	\hat{ \vec{x}}^{(t)} = \left(  \mtx{H}^{T}\mtx{H} + \frac{\beta}{2} \mtx{I}  \right)^{-1} \left( \mtx{H}^T \vec{y} + \frac{\beta}{2} \sum\limits_{i\in \mathcal{B}}{ \mtx{P}_i^T {\tilde{\vec{z}}}_{i}^{(t)} }  \right) \nonumber\\
\end{IEEEeqnarray}
rendering this stage as a weighted averaging of the deteriorated signal with the block estimates obtained in the second stage of the previous iteration.
While the analytic solution (\ref{eq:complexity-regularized restoration - iterative solution - inversion - analytic form}) explains the underlying meaning of the $ \ell_2 $-constrained deconvolution stage (\ref{eq:complexity-regularized restoration - iterative solution - inversion}), it includes matrix inversion that, in general, may lead to numerical instabilities. Accordingly, in the implementation of the proposed method we suggest to address (\ref{eq:complexity-regularized restoration - iterative solution - inversion}) via numerical optimization techniques (for example, we used the biconjugate gradients method).

The optimizations in the second stage of each iteration (\ref{eq:complexity-regularized restoration - iterative solution - compression}) are rate-distortion optimizations corresponding to each of the non-overlapping blocks of the signal estimate $ \hat{ \vec{x}}^{(t)} $ obtained in the first stage. Accordingly, the set of block-level optimizations in (\ref{eq:complexity-regularized restoration - iterative solution - compression}) can be interpreted as a single full-signal rate-distortion optimization with respect to a Lagrange multiplier value of $ {\lambda} = \frac{2\mu}{\beta} $.
We denote the compression-decompression procedure that replaces (\ref{eq:complexity-regularized restoration - iterative solution - compression}) as 
\begin{IEEEeqnarray}{rCl}
	\label{eq:compression-decompression function}
	\hat{\vec{z}}^{(t)}=CompressDecompress_{\lambda} \left( \tilde{\vec{x}}^{(t)}  \right),
\end{IEEEeqnarray}
where $  \tilde{\vec{x}}^{(t)} \triangleq  \mathop \sum\limits_{i\in\mathcal{B}} \mtx{P}_i^T \tilde{\vec{x}}_i^{(t)} $ is the signal to compress, assembled from all the non-overlapping blocks, and $ \hat{\vec{z}}^{(t)} $ is the corresponding decompressed full signal. 
Moreover, by defining a full-sized scaled dual variable ${\vec{u}}^{(t)} \triangleq  \mathop \sum\limits_{i\in\mathcal{B}} \mtx{P}_i^T{\vec{u}}_i^{(t)} $ we get that $  \tilde{\vec{x}}^{(t)} =  \hat{\vec{x}}^{(t)} + {\vec{u}}^{(t)}$. Then, using the definitions established here we can translate the block-level computations (\ref{eq:complexity-regularized restoration - iterative solution - inversion})-(\ref{eq:complexity-regularized restoration - iterative solution - beta update}) into the full-signal formulations described in Algorithm \ref{Algorithm:Proposed Method Non-overlapping}.

We further suggest using a standardized compression method as the compression-decompression operator (\ref{eq:compression-decompression function}). While many compression methods do not follow the exact rate-distortion optimizations we have in our mathematical development, we still encourage utilizing such techniques as an approximation for (\ref{eq:complexity-regularized restoration - iterative solution - compression}). 
Additionally, since many compression methods do not rely on Lagrangian optimization, their operating parameters may have different definitions such as quality parameters, compression ratios, or output bit-rates. 
Accordingly, we present the suggested algorithm with respect to a general compression-decompression procedure with output bit-cost directly or indirectly affected by a parameter denoted as $ \theta $. 
These generalizations are also implemented in the proposed Algorithm \ref{Algorithm:Proposed Method Non-overlapping}.
In Section \ref{sec:Experimental Results} we elaborate on particular settings of $ \theta $ that were empirically found appropriate for utilization of the HEVC and the JPEG2000 standard. In cases where the compression method significantly deviates from a Lagrangian optimization form, it can be useful to appropriately update the compression parameter in each iteration (this is the case for JPEG2000 as explained in Section \ref{sec:Experimental Results}).

Importantly, Algorithm \ref{Algorithm:Proposed Method Non-overlapping} does not only restore the deteriorated input image, but also provides the signal estimate in a compressed form by employing the output of the compression stage of the last iteration.

\begin{algorithm}
	\caption{Proposed Method Based on Total Complexity of Non-Overlapping Blocks}\label{Algorithm:Proposed Method Non-overlapping}
	\begin{algorithmic}[1]
		\State  Inputs: $\vec{y}$, $ \beta $, $ \theta $.
		\State  Initialize $ {\hat{\vec{z}}}^{(0)} $ (depending on the deterioration type).
		\State $t = 1$ and $ {\vec{u}}^{(1)} = \vec{0} $
		\Repeat
		
		\State $\tilde{\vec{z}}^{(t)} =  {\hat{\vec{z}}}^{(t-1)} - \vec{u}^{(t)}$
		\State Solve the $ \ell_2 $-constrained deconvolution:\newline $~~~~~~~~\hat{ \vec{x}}^{(t)} = \mathop {\text{argmin}}\limits_{\vec{x}} \left\| {  \mtx{H} \vec{x} - \vec{y} } \right\|_2^2 + \frac{\beta}{2}  {\left\| {   \vec{x} - \tilde{\vec{z}}^{(t)} } \right\|_2^2} $
		
		\State $\tilde{\vec{x}}^{(t)} =  {\hat{\vec{x}}}^{(t)} + \vec{u}^{(t)}$
		\State $ \hat{\vec{z}}^{(t)} = {CompressDecompress}_{\theta}\left( \tilde{\vec{x}}^{(t)} \right) $
		\vspace{0.05in}

		\State $\vec{u}^{(t+1)} = \vec{u}^{(t)} + \left( \hat{\vec{x}}^{(t)} - \hat{\vec{z}}^{(t)} \right)$
		\State $ t \gets t + 1$
		\Until{stopping criterion is satisfied}
	\end{algorithmic}
\end{algorithm}

\subsection{Regularize Total Complexity of All Overlapping Blocks}
\label{subsec:Complexity of Overlapping Blocks}

Algorithm \ref{Algorithm:Proposed Method Non-overlapping} emerged from complexity regularization measuring the total bit-cost of the estimate based on its decomposition into non-overlapping blocks (see Eq. (\ref{eq:complexity-regularized restoration - expressing blocks})), resulting in a restored signal available in a compressed form compatible with the compression technique in use.
Obviously, the above approach provides estimates limited to the discrete set of signals supported by the compression architecture, thus, having a somewhat reduced restoration ability with respect to methods providing estimates from an unrestricted domain of solutions. 
This observation motivates us to develop a complexity-regularized restoration procedure that provides good estimates from the continuous unrestricted domain of signals while still utilizing a standardized compression technique as its main component.

As before, our developments refer to a general block-based compression method relying on a codebook $ \mathcal{C} $ as a discrete set of block reconstruction candidates. 
We consider here the segmentation of the signal-block space, $ \mathbb{R}^{N_b} $, given by the voronoi cells corresponding to the compression reconstruction candidates, namely, for each $ \vec{c} \in \mathcal{C} $ there is a region 
\begin{IEEEeqnarray}{rCl}
	\label{eq:voronoi cell of codebook element}
	V_{\vec{c}} \triangleq \left\lbrace  \vec{w} \in \mathbb{R}^{N_b} ~~ \Big|{~~  \vec{c} = \arg\min_{\tilde{\vec{c}}\in\mathcal{C}} \left\| \vec{w} - \tilde{\vec{c}} \right\|_2^2   } \right\rbrace 
\end{IEEEeqnarray}
defining all the vectors in $ \mathbb{R}^{N_b} $ that $ \vec{c} $ is their nearest member of $ \mathcal{C} $.
We use the voronoi cells in (\ref{eq:voronoi cell of codebook element}) for defining an alternative extension to the bit-cost evaluation of a signal block (i.e., the new definition, $\bar{r}_{v}(\vec{z})$, will replace $\bar r(\vec{z})$ given in (\ref{eq:extended block bit-cost}) that was used for the development of Algorithm \ref{Algorithm:Proposed Method Non-overlapping}). Specifically, we associate a finite bit-cost to any $ \vec{z} \in \mathbb{R}^{N_b} $ based on the voronoi cell it resides in, i.e., 
\begin{IEEEeqnarray}{rCl}
	\label{eq:voronoi extended block bit-cost}
	\bar{r}_{v}(\vec{z}) = r(\vec{c}) \text{~~for~~}\vec{z}\in V_{\vec{c}}
\end{IEEEeqnarray}
where $ r(\vec{c}) $ is the regular bit-cost evaluation defined in Section \ref{subsec:Operational Rate-Distortion Optimization} only for blocks in $\mathcal{C}$.

The method proposed here emerges from a new complexity regularization function that quantifies the total complexity of all the overlapping blocks of the estimate. Using the extended bit-cost measure $ \bar{r}_{v}(\cdot) $, defined in (\ref{eq:voronoi extended block bit-cost}), we introduce the full-signal regularizer as 
\begin{IEEEeqnarray}{rCl}
	\label{eq:complexity regularizer of all overlapping blocks}
	s^{*}(\vec{x})\nolinebreak=\nolinebreak \sum\limits_{i\in \mathcal{B}^*} \bar {r}_v (\mtx{P}_i\vec{x})
\end{IEEEeqnarray}
where $ \vec{x} \in \mathbb{R}^N $, and $ \mathcal{B}^* $ is a set containing the indices of all the overlapping blocks of the signal. 
The associated restoration optimization is 
\begin{IEEEeqnarray}{rCl}
	\label{eq:complexity-regularized restoration - overlapping blocks}
	\hat{ \vec{x}} = \mathop {{\text{argmin}}}\limits_{\vec{x}} \left\| {  \mtx{H} \vec{x} - \vec{y} } \right\|_2^2 + \mu  \sum\limits_{i\in \mathcal{B}^*} \bar{r}_v (\mtx{P}_i\vec{x}).
\end{IEEEeqnarray}
Importantly, in contrast to the previous subsection, the function $ s^{*}(\vec{x}) $ evaluates the complexity of any $ \vec{x} \in \mathbb{R}^N $ with a finite value and, thus, does not restrict the restoration to the discrete set of codebook-based constructions, $ \mathcal{C}_{\mathcal{B}} $, defined in (\ref{eq:group of solutions based on nonoverlapping blocks}). 

While the new regularizer in (\ref{eq:complexity-regularized restoration - overlapping blocks}) is not separable into complexity evaluation of non-overlapping blocks, the ADMM approach can accommodate it as well. This is explained next. 
We define the auxiliary variables $ \left\lbrace \vec{z}_i \right\rbrace_{i\in\mathcal{B}^*} $, where each $ \vec{z}_i $ is coupled with the $ i^{th} $ overlapping block. Then, the optimization (\ref{eq:complexity-regularized restoration - overlapping blocks}) is expressed as 
\begin{IEEEeqnarray}{rCl}
	\label{eq:complexity-regularized restoration - overlapping blocks - with splitting - constrained form}
	&&\left( \hat{ \vec{x}}, \left\lbrace \hat{\vec{z}}_i \right\rbrace_{i\in\mathcal{B}^*} \right) = \mathop {{\text{argmin}}}\limits_{\vec{x},\left\lbrace \vec{z}_i \right\rbrace_{i\in\mathcal{B}^*}} \left\| {  \mtx{H} \vec{x} - \vec{y} } \right\|_2^2 + 
	\mu  \mathop\sum\limits_{i\in \mathcal{B}^*} {\bar{r}_v (\vec{z}_i)} \nonumber\\ 
	&& \text{s.t. }~~~~~ \vec{z}_i = \mtx{P}_i \vec{x} ~\text{,~~for~} i\in\mathcal{B}^* . 
\end{IEEEeqnarray}
As in Section \ref{subsec:Total Complexity of Non-Overlapping Blocks}, employing the augmented Lagrangian (in its scaled form) and the method of multipliers results in an iterative solution provided by the following three steps in each iteration (as before $ t $ denotes the iteration number):
\begin{IEEEeqnarray}{rCl}
	\label{eq:complexity-regularized restoration - overlapping blocks - iterative solution - inversion}
	&&  \hat{ \vec{x}}^{(t)} = \mathop {\text{argmin}}\limits_{\vec{x}} \left\| {  \mtx{H} \vec{x} - \vec{y} } \right\|_2^2 + \frac{\beta}{2} \sum\limits_{i\in \mathcal{B}^{*}} {\left\| {  \mtx{P}_i \vec{x} - {\tilde{\vec{z}}}_{i}^{(t)}} \right\|_2^2}~~~~   \\ 
	\label{eq:complexity-regularized restoration - overlapping blocks - iterative solution - compression}
	&&  \hat{\vec{z}}_i^{(t)} = \mathop {\text{argmin}}\limits_{\vec{z}_i} \frac{\beta}{2}   {\left\| {  {\tilde{\vec{x}}}_{i}^{(t)} - \vec{z}_i } \right\|_2^2}   + \mu   \bar r(\vec{z}_i),   ~~ i\in\mathcal{B}^{*} \\
	\label{complexity-regularized restoration - overlapping blocks - iterative solution - beta update}
	&& 	\vec{u}_i^{(t+1)} = \vec{u}_i^{(t)} + \left( \mtx{P}_i \hat{\vec{x}}^{(t)} - \hat{\vec{z}}_i^{(t)} \right)	,   ~~ i\in\mathcal{B}^{*} , 
\end{IEEEeqnarray}
where $ \vec{u}_i^{(t)} $ is the scaled dual variable for the $ i^{th} $ block, $ {\tilde{\vec{z}}}_{i}^{(t)} \nolinebreak \triangleq \nolinebreak  {\hat{\vec{z}}}_{i}^{(t-1)} - \vec{u}_i^{(t) } $ and $ {\tilde{\vec{x}}}_{i}^{(t)} \triangleq \mtx{P}_i \hat{\vec{x}}^{(t)} + \vec{u}_i^{(t)} $  for $  i\in\mathcal{B}^{*} $.

While the procedure above resembles the one from the former subsection, the treatment of overlapping blocks has different interpretations to the optimizations in (\ref{eq:complexity-regularized restoration - overlapping blocks - iterative solution - inversion}) and (\ref{eq:complexity-regularized restoration - overlapping blocks - iterative solution - compression}).
Indeed, note that the block-level rate-distortion optimizations in (\ref{eq:complexity-regularized restoration - overlapping blocks - iterative solution - compression}) are not discrete due to the extended bit-cost evaluation $ \bar{r}_v(\cdot) $ defined in (\ref{eq:voronoi extended block bit-cost}). Due to the definition of $ \bar{r}_v(\cdot) $, the rate-distortion optimizations (\ref{eq:complexity-regularized restoration - overlapping blocks - iterative solution - compression}) can be considered as continuous relaxations of the discrete optimizations done by the practical compression technique. Since we intend using a given compression method without explicit knowledge of its underlying codebook, we cannot construct the voronoi cells defining $ \bar{r}_v(\cdot) $ and, thus, it is impractical to accurately solve  (\ref{eq:complexity-regularized restoration - overlapping blocks - iterative solution - compression}). Consequently, we suggest to  approximate the optimizations (\ref{eq:complexity-regularized restoration - overlapping blocks - iterative solution - compression}) by the discrete forms of 
\begin{IEEEeqnarray}{rCl}
	\label{eq:complexity-regularized restoration - overlapping blocks - iterative solution - compression - discrete}
	&&  \hat{\vec{z}}_i^{(t)} = \mathop {\text{argmin}}\limits_{\vec{z}_i} \frac{\beta}{2}   {\left\| {  \mtx{P}_i \hat{\vec{x}}^{(t)} - \vec{z}_i } \right\|_2^2}   + \mu   \bar{r}(\vec{z}_i),   ~ i\in\mathcal{B}^* 
\end{IEEEeqnarray}
where $ \bar{r}(\cdot) $ is the discrete evaluation of the block bit-cost, defined in (\ref{eq:extended block bit-cost}), letting to identify the problems as operational rate-distortion optimizations of the regular discrete form.

Each block-level rate-distortion optimization in (\ref{eq:complexity-regularized restoration - overlapping blocks - iterative solution - compression - discrete}) is associated with one of the overlapping blocks of the signal. 
Accordingly, we interpret this group of optimizations as multiple applications of a full-signal compression-decompression procedure, each
associates to a shifted version of the signal (corresponding to different sets of non-overlapping blocks).
Specifically, for a signal $ \vec{x} $ and a compression block-size of $ N_b $ samples, there are $ N_b $ shifted grids of non-overlapping blocks. 
For mathematical convenience, we consider here cyclic shifts such that the $ j^{th} $ shift ($ j=1,...,N_b $) corresponds to a signal of $ N $ samples taken cyclically starting at the $ j^{th} $ sample of $ \vec{x} $ (in practice other definitions of shifts may be used, e.g., see Section \ref{sec:Experimental Results} for a suggested treatment of two-dimensional signals). We denote the $ j^{th} $ shifted signal as $ shift_{j}\left\lbrace \vec{x} \right\rbrace $.
Moreover, we denote the index set of blocks included in the $ j^{th} $ shifted signal as $ \mathcal{B}^{j} $ (noting that  $ \mathcal{B}^{1} =  \mathcal{B} $), hence, $ \mathcal{B}^{*} = \cup_{j=1}^{N_{b}}\mathcal{B}^{j} $.
Therefore, the decompressed blocks $\left\lbrace{\hat{\vec{z}}_i^{(t)}}\right\rbrace_{i\in\mathcal{B}^*}$ can be decomposed into $ N_{b} $ subsets, $\left\lbrace{\hat{\vec{z}}_i^{(t)}}\right\rbrace_{i\in\mathcal{B}^j}$ for $ j=1,...,N_b $, each contains non-overlapping blocks corresponding to a different shifted grid. Moreover, the $ j^{th} $ set of blocks, $\left\lbrace{\hat{\vec{z}}_i^{(t)}}\right\rbrace_{i\in\mathcal{B}^j}$, is associated with the full signal $ \hat{\vec{z}}^{j,(t)} \triangleq \sum\limits_{i\in \mathcal{B}^j}{ \mtx{P}_i^T  \hat{\vec{z}}_{i}^{(t)}}  $. Then, the set of full signals $\left\lbrace{\hat{\vec{z}}^{j,(t)}}\right\rbrace_{j=1}^{N_b}$ can be obtained by multiple full-signal compression-decompression applications, namely, for $ j\nolinebreak=\nolinebreak 1,...,N_b $: $ \hat{\vec{z}}_{shifted}^{j,(t)}\nolinebreak=\nolinebreak CompressDecompress_{ \lambda} \left( \tilde{\vec{x}}^{j,(t)}_{shifted} \right) $, where the Lagrangian multiplier value is $ {\lambda} = \frac{2\mu}{\beta} $ and 
\begin{IEEEeqnarray}{rCl}
	\label{eq:complexity-regularized restoration - overlapping blocks - shifted compression input - definition}
	&&   \tilde{\vec{x}}^{j,(t)}_{shifted} \triangleq  shift_{j}\left\lbrace \hat{\vec{x}}^{(t)}  + {\vec{u}}^{j,(t)} \right\rbrace
\end{IEEEeqnarray}
is the compression input formed as the $ j^{th} $ shift of $ \hat{\vec{x}}^{(t)} $ combined with the full-sized dual variable defined via 
\begin{IEEEeqnarray}{rCl}
	\label{eq:complexity-regularized restoration - overlapping blocks - shifted dual variable - definition}
	&&   {\vec{u}}^{j,(t)} \triangleq  \sum\limits_{i\in \mathcal{B}^j}{ \mtx{P}_i^T {\vec{u}}_{i}^{(t)}}
\end{IEEEeqnarray}
assembled from the block-level dual variables corresponding to the $ j^{th} $ grid of non-overlapping blocks. Notice that inverse shifts are required for obtaining the desired signals, i.e., 
\begin{IEEEeqnarray}{rCl}
	\label{eq:complexity-regularized restoration - overlapping blocks - shifted decompression output - definition}
	&&    \hat{\vec{z}}^{j,(t)} = shift_{j}^{-1}\left\lbrace \hat{\vec{z}}_{shifted}^{j,(t)} \right\rbrace
\end{IEEEeqnarray}
where $ shift^{-1}_{j} \left\lbrace \cdot \right\rbrace$ is the inverse shift operator that (cyclically) shifts back the given full-size signal by $ j $ samples.

The deconvolution stage (\ref{eq:complexity-regularized restoration - overlapping blocks - iterative solution - inversion}) of the iterative process can be rewritten as 
\begin{IEEEeqnarray}{rCl}
	\label{eq:complexity-regularized restoration - overlapping blocks - iterative solution - inversion - full signals}
	&&  \hat{ \vec{x}}^{(t)} = \mathop {\text{argmin}}\limits_{\vec{x}} \left\| {  \mtx{H} \vec{x} - \vec{y} } \right\|_2^2 + \frac{\beta}{2} \sum\limits_{j=1}^{N_b} {\left\| {  \vec{x} - {\tilde{\vec{z}}}^{j,(t)}} \right\|_2^2}~~~~  
\end{IEEEeqnarray}
where the regularization part (the second term) considers the distance of the estimate from the $ N_b $ full signals defined via 
\begin{IEEEeqnarray}{rCl}
	\label{eq:complexity-regularized restoration - overlapping blocks - iterative solution - inversion - full signals - auxiliary definition}
	&&  {\tilde{\vec{z}}}^{j,(t)}  \triangleq  {\hat{\vec{z}}}^{j,(t)} - {\vec{u}}^{j,(t)}
\end{IEEEeqnarray}
for $ j=1,...,N_b $, where $ {\hat{\vec{z}}}^{j,(t)} $ and $ {\vec{u}}^{j,(t)} $ were defined above.
The analytic solution of the optimization (\ref{eq:complexity-regularized restoration - overlapping blocks - iterative solution - inversion - full signals}) is 
\begin{IEEEeqnarray}{rCl}
	\label{eq:complexity-regularized restoration - overlapping blocks - inversion - analytic form - full signals}
	&&\hat{ \vec{x}}^{(t)} = \left(  \mtx{H}^{T}\mtx{H} + \frac{\beta}{2} N_b \right)^{-1} \left( \mtx{H}^T \vec{y} + \frac{\beta}{2} \sum\limits_{j=1}^{N_b}{ {\tilde{\vec{z}}}^{j,(t)} }  \right) , ~~
\end{IEEEeqnarray}
showing that the first stage of each iteration is a weighted averaging of the given deteriorated signal with all the decompressed signals (and the dual variables) obtained in the former iteration.
It should be noted that the analytic solution (\ref{eq:complexity-regularized restoration - overlapping blocks - inversion - analytic form - full signals}) is developed here for showing the essence of the $ \ell_2 $-constrained deconvolution part of the method. Nevertheless, the possible numerical instabilities due to the matrix inversion appearing in (\ref{eq:complexity-regularized restoration - overlapping blocks - inversion - analytic form - full signals}) motivate the practical direct treatment of (\ref{eq:complexity-regularized restoration - overlapping blocks - iterative solution - inversion}) via numerical optimization techniques.

Algorithm \ref{Algorithm:Proposed Method Overlapping Blocks} summarizes the practical restoration method for a compression technique operated by the general parameter $ {\theta} $ for determining the bit-cost (see details in Section \ref{subsec:Total Complexity of Non-Overlapping Blocks}).

The computational cost of Algorithm \ref{Algorithm:Proposed Method Overlapping Blocks} stems from its reliance on repeated applications of compressions, decompressions, and $ \ell_2 $ - constrained deconvolution procedures. While the actual run-time depends on the computational complexity of the utilized compression technique, we can generally state that the total run-time will be of at least the run-time of compression and decompression processes for a total number of applications equal to the product of the number of iterations and the number of shifts considered.

\begin{algorithm}
	\caption{Proposed Method Based on Total Complexity of All the Overlapping Blocks}\label{Algorithm:Proposed Method Overlapping Blocks}
	\begin{algorithmic}[1]
		\State  Inputs: $ \vec{y} $, $\beta$, $ \theta $.
		\State  Initialize $ \left\lbrace {\hat{\vec{z}}}^{j,(0)} \right\rbrace_{j=1}^{N_b} $ (depending on the deterioration type).
		\State $t = 1$ and  $ {{\vec{u}}}^{j,(1)} = \vec{0} $ for $ j = 1,...,{N_b} $.
		\Repeat
		
		\State ${\tilde{\vec{z}}}^{j,(t)} = {\hat{\vec{z}}}^{j,(t-1)} - \vec{u}^{j,(t) }  \text{~~for~}j=1,...,N_b  $
		\State Solve the $ \ell_2 $-constrained deconvolution:\newline $~~~~\hat{ \vec{x}}^{(t)} = \mathop {\text{argmin}}\limits_{\vec{x}} \left\| {  \mtx{H} \vec{x} - \vec{y} } \right\|_2^2 + \frac{\beta}{2} \sum\limits_{j=1}^{N_b} {\left\| {  \vec{x} - {\tilde{\vec{z}}}^{j,(t)}} \right\|_2^2} $
		
		\For{$ j=1,...,N_b $}
		\State $\tilde{\vec{x}}_{shifted}^{j,(t)} = shift_{j}\left\lbrace \hat{\vec{x}}^{(t)}  + {\vec{u}}^{j,(t)} \right\rbrace$
		
		\State $ \hat{\vec{z}}_{shifted}^{j,(t)} = {CompressDecompress}_{\theta}\left( \tilde{\vec{x}}_{shifted}^{j,(t)}  \right) $
		
		\State $\hat{\vec{z}}^{j,(t)} = shift^{-1}_{j}\left\lbrace { \hat{\vec{z}}_{shifted}^{j,(t)} } \right\rbrace$
		
		\State $\vec{u}^{j,(t+1)} = \vec{u}^{j,(t)} + \left( \hat{\vec{x}}^{(t)} - \hat{\vec{z}}^{j,(t)} \right) $
		
		\EndFor

		\State $ t \gets t + 1$
		\Until{stopping criterion is satisfied}
	\end{algorithmic}
\end{algorithm}

The ADMM is known for promoting distributed optimization structures \cite{boyd2011distributed}. In Algorithm \ref{Algorithm:Proposed Method Overlapping Blocks} the distributed nature of the ADMM is expressed in the separate optimization of each of the shifted block-grids (see stages 8-11). In particular, the dual variables $ \left\lbrace \vec{u}^{j,(t)} \right\rbrace_{j=1}^{N_b} $, associated with the various grids (see stages 5,8, and 11 in Algorithm \ref{Algorithm:Proposed Method Overlapping Blocks}), are updated independently in stage 11 such that each considers only its respective $ \hat{\vec{z}}^{j,(t)} $. However, the dual variables $ \left\lbrace \vec{u}^{j,(t)} \right\rbrace_{j=1}^{N_b} $  essentially refer to the same data based on different block-grids. Accordingly, we suggest to merge the independent dual variables to form a single, more robust, dual variable defined as 
\begin{IEEEeqnarray}{rCl}
	\label{eq:complexity-regularized restoration - overlapping blocks - robust dual variable}
	&& \vec{u}_{total}^{(t)} = \frac{1}{N_b} \sum\limits_{j=1}^{N_b}{ \vec{u}^{j,(t)}  } 
\end{IEEEeqnarray}
where the averaging tends to reduce particular artifacts that may appear due to specific block-grids.
We utilize the averaged dual variable (\ref{eq:complexity-regularized restoration - overlapping blocks - robust dual variable}) to extend Algorithm \ref{Algorithm:Proposed Method Overlapping Blocks} into Algorithm \ref{Algorithm:Proposed Method Overlapping Blocks with Robust Dual Variables}. Notice stages 5,8, and 13 of Algorithm \ref{Algorithm:Proposed Method Overlapping Blocks with Robust Dual Variables}, where the averaged dual variable is used instead of the independent ones. 
\begin{algorithm}
	\caption{Proposed Method Based on Total Complexity of All the Overlapping Blocks with Robust Dual Variables}\label{Algorithm:Proposed Method Overlapping Blocks with Robust Dual Variables}
	\begin{algorithmic}[1]
		\State  Inputs: $ \vec{y} $, $\beta$, $ \theta $.
		\State  Initialize $ \left\lbrace {\hat{\vec{z}}}^{j,(0)} \right\rbrace_{j=1}^{N_b} $ (depending on the deterioration type).
		\State $t = 1$ and  $ \vec{u}_{total}^{(1)} = \vec{0} $.
		\Repeat
		
		\State ${\tilde{\vec{z}}}^{j,(t)} = {\hat{\vec{z}}}^{j,(t-1)} - \vec{u}_{total}^{(t)}  \text{~~for~}j=1,...,N_b  $
		\State Solve the $ \ell_2 $-constrained deconvolution:\newline $~~~~\hat{ \vec{x}}^{(t)} = \mathop {\text{argmin}}\limits_{\vec{x}} \left\| {  \mtx{H} \vec{x} - \vec{y} } \right\|_2^2 + \frac{\beta}{2} \sum\limits_{j=1}^{N_b} {\left\| {  \vec{x} - {\tilde{\vec{z}}}^{j,(t)}} \right\|_2^2} $
		
		\For{$ j=1,...,N_b $}
		\State $\tilde{\vec{x}}_{shifted}^{j,(t)} = shift_{j}\left\lbrace \hat{\vec{x}}^{(t)}  + \vec{u}_{total}^{(t)} \right\rbrace$
		
		\State $ \hat{\vec{z}}_{shifted}^{j,(t)} = {CompressDecompress}_{\theta}\left( \tilde{\vec{x}}_{shifted}^{j,(t)}  \right) $
		
		\State $\hat{\vec{z}}^{j,(t)} = shift^{-1}_{j}\left\lbrace { \hat{\vec{z}}_{shifted}^{j,(t)} } \right\rbrace$
		
		\State $\vec{u}^{j,(t+1)} = \vec{u}^{j,(t)} + \left( \hat{\vec{x}}^{(t)} - \hat{\vec{z}}^{j,(t)} \right) $
		
		\EndFor
		
		\State $\vec{u}_{total}^{(t+1)} = \frac{1}{N_b} \sum\limits_{j=1}^{N_b}{ \vec{u}^{j,(t+1)}  }$
		
		\State $ t \gets t + 1$
		\Until{stopping criterion is satisfied}
	\end{algorithmic}
\end{algorithm}

In Section \ref{sec:Experimental Results} we further discuss practical aspects of the proposed Algorithms \ref{Algorithm:Proposed Method Non-overlapping}-\ref{Algorithm:Proposed Method Overlapping Blocks with Robust Dual Variables} and evaluate their performance for deblurring and inpainting of images.

\section{Rate-Distortion Theoretic Analysis for the Gaussian Case}
\label{sec:Rate-Distortion Theoretic Analysis for the Gaussian Case}
In this section we theoretically study the complexity-regularized restoration problem from the perspective of rate-distortion theory. While our analysis is focused on the particular settings of a cyclo-stationary Gaussian signal and deterioration caused by a linear shift-invariant operator and additive white Gaussian noise, the results clearly explain the main principles of complexity-regularized restoration.

In general, theoretical studies of rate-distortion problems for the Gaussian case provide to the signal processing practice optimistic beliefs about which design concepts perform well for the real-world non-Gaussian instances of the problems (see the excellent discussion in \cite[Sec. 3]{donoho1998data}). 
Moreover, theoretical and practical solutions may embody in a different way the same general concepts. Therefore, one should look for connections between theory and practice in the form of high-level analogies.

The optimal solution presented in this section considers the classical framework of rate-distortion theory and a particular, however, important case of a Gaussian signal and a linear shift-invariant degradation operator. 
Our rate-distortion analysis below will show that the optimal complexity-regularized restoration consists of the following two main ideas: pseudoinverse filtering of the degraded input, and compression with respect to a squared-error metric that is weighted based on the degradation-filter squared-magnitude (considering a processing in the Discrete Fourier Transform (DFT) domain). 
In Subsection \ref{subsec:Rate-Distortion Theoretic - Conceptual Relation to The Proposed Approach} we explain how these two concepts connect to more general themes having different realizations in the practical approach proposed in Section \ref{sec:Proposed Methods}.

In this section, consider the signal  $ \vec{x} \in \mathbb{R}^N $ modeled as a zero-mean cyclo-stationary Gaussian random vector with a circulant autocorrelation matrix $ \mtx{R}_{\vec{x}} $, i.e.,  $\vec{x} \sim  \mathcal{N}\left(0,\mtx{R}_{\vec{x}}\right) $.
The degradation model studied remains
\begin{IEEEeqnarray}{rCl}
	\label{eq:theoretic Gaussian analysis - degradation model}
	{\vec{y}} = \mtx{H} \vec{x} + \vec{n} , 
\end{IEEEeqnarray}
where here $ \mtx{H} $ is a real-valued $N\times N$ circulant  matrix representing a linear shift-invariant deteriorating operation and $\vec{n} \sim  \mathcal{N}\left(0,\sigma_n^2\mtx{I}\right) $ is a length $N$ vector of white Gaussian noise.
Clearly, the degraded observation $ \vec{y} $ is also a zero-mean cyclo-stationary Gaussian random vector with a circulant autocorrelation matrix $ \mtx{R}_{\vec{y}}\nolinebreak=\nolinebreak\mtx{H}  \mtx{R}_{\vec{x}} \mtx{H}^* + \sigma_n^2\mtx{I} $.

\subsection{Prevalent Restoration Strategies}
\label{subsec:Alternative Restoration Strategies}

We precede the analysis of the complexity-regularized restoration with mentioning three well-known estimation methods.
The restoration procedure is a function 
\begin{IEEEeqnarray}{rCl}
	\label{eq:theoretic Gaussian analysis - restoration function}
	{\hat{\vec{x}}} = f \left( \vec{y} \right), 
\end{IEEEeqnarray}
where $ f $ maps the degraded signal $ \vec{y} $ to an estimate of $ \vec{x} $ denoted as $ \hat{\vec{x}} $. In practice, one gets a realization of $ \vec{y} $ denoted here as $ \vec{y}_r $ and forms the corresponding estimate as $ {\hat{\vec{x}}}_r \nolinebreak = \nolinebreak f \left( \vec{y}_r \right)  $.

\subsubsection{Minimum Mean Squared Error (MMSE) Estimate}
This restoration minimizes the expected MSE of the estimate, i.e., 
\begin{IEEEeqnarray}{rCl}
	\label{eq:theoretic Gaussian analysis - MMSE estimate optimization}
	 f_{MMSE} = \underset{ f  }{\text{argmin}} ~ E \left\lbrace \left\| { \vec{x} - f \left( \vec{y} \right) } \right\| _2^2 \right\rbrace , 
\end{IEEEeqnarray}
yielding that the corresponding estimate is the conditional expectation of $ \vec{x} $ given $ \vec{y} $ 
\begin{IEEEeqnarray}{rCl}
	\label{eq:theoretic Gaussian analysis - MMSE estimate}
	{\hat{\vec{x}}}_{MMSE} = f_{MMSE} \left( \vec{y} \right) = E\left\lbrace \vec{x} \vert \vec{y} \right\rbrace .
\end{IEEEeqnarray}
Nicely, for the Gaussian case considered in this section, the MMSE estimate (\ref{eq:theoretic Gaussian analysis - MMSE estimate}) reduces to a linear operator, presented below as the Wiener filter.

\subsubsection{Wiener Filtering}

The Wiener filter is also known as the Linear Minimum Mean Squared Error (LMMSE) estimate, corresponding to a restoration function of the form 
\begin{IEEEeqnarray}{rCl}
	\label{eq:theoretic Gaussian analysis - Wiener restoration function}
	{\hat{\vec{x}}} = f_{Wiener} \left( \vec{y} \right) = \mtx{A}\vec{y} + \vec{b} , 
\end{IEEEeqnarray}
optimized via  
\begin{IEEEeqnarray}{rCl}
	\label{eq:theoretic Gaussian analysis - Wiener estimate optimization}
	\left\lbrace {\hat{\mtx{A}},\hat{\vec{b}}} \right\rbrace = \underset{ \mtx{A},\vec{b}  }{\text{argmin}} ~ E \left\lbrace \left\| {  \vec{x} - \left( \mtx{A}\vec{y} + \vec{b} \right) }\right\| _2^2 \right\rbrace . 
\end{IEEEeqnarray}
In our case, where $ \vec{x} $ and $ \vec{y} $ are zero mean, $ \hat{\vec{b}} = \vec{0} $ and 
\begin{IEEEeqnarray}{rCl}
	\label{eq:theoretic Gaussian analysis - Wiener filter matrix}
	\hat{\mtx{A}} = \mtx{R}_{\vec{x}} \mtx{H}^* \left(  \mtx{H}  \mtx{R}_{\vec{x}} \mtx{H}^* + \sigma_n^2\mtx{I}  \right)^{-1} .
\end{IEEEeqnarray}	
If the distributions are Gaussian, this linear operator coincides with the optimal MMSE estimator.

\subsubsection{Constrained Deconvolution Filtering}

This approach considers a given degraded signal $ \vec{y}_r = \mtx{H} \vec{x}_0 + \vec{n}_r $, with the noise vector a realization of a random process while the signal $\vec{x}_0$ is considered as a deterministic vector, with perhaps some known properties. 
Then, the restoration is carried out by minimizing a carefully-designed penalty function, $ g $, that assumes lower values for $\vec{x}$ vectors that fit the prior knowledge on $\vec{x}_0$. 
Note that for a sufficiently large signal dimension we get that $ \left\| { \vec{y}_r  - \mtx{H} \vec{x}_0 } \right\|_2^2  = \left\| \vec{n}_r \right\|_2^2 \approx N \sigma_n^2 $. The last result motivates to constrain the estimate, $ \hat{\vec{x}} $, to conform with the known degradation model (\ref{eq:theoretic Gaussian analysis - degradation model}), by demanding the similarity of $  { \vec{y}_r  - \mtx{H} \hat{\vec{x}} } $ to the additive noise term via the equality relation $ \left\| { \vec{y}_r  - \mtx{H} \hat{\vec{x}} } \right\|_2^2  = N \sigma_n^2 $. The above idea is implemented in an optimization of the form 
\begin{IEEEeqnarray}{rCl}
	\label{eq:theoretic Gaussian analysis - constrained deconvolution filtering}
	\begin{aligned}
		& \underset{ \hat{\vec{x}} } {\text{min}}
		& & { g\left( \hat{\vec{x}} \right) } \\
		& \text{subject to}
		& & \left\| { \vec{y}_r  - \mtx{H} \hat{\vec{x}} } \right\|_2^2  =  N \sigma_n^2 .
	\end{aligned}	
\end{IEEEeqnarray}
Our practical methods presented in Section \ref{sec:Proposed Methods} emerge from an instance of the constrained deconvolution optimization (\ref{eq:theoretic Gaussian analysis - constrained deconvolution filtering}), in its Lagrangian version, where the penalty function $ g $ is the cost in bits measuring the complexity in describing the estimate $ \hat{\vec{x}} $.
In the remainder of this section, we study the complexity-regularized restoration problem from a statistical perspective.

\subsection{The Complexity-Regularized Restoration Problem and its Equivalent Forms}
\label{subsec:Reformulations of the Problem}

Based on rate-distortion theory (e.g., see \cite{cover2012elements}), we consider the estimate of $ {\vec{x}} $ as a random vector $ \hat{\vec{x}} \in \mathbb{R}^N $ with the probability density function (PDF) $ p_{\hat{\vec{x}}}\left(  \vecRvReal{\hat{x}} \right) $. The estimate characterization, $ p_{\hat{\vec{x}}}\left(  \vecRvReal{\hat{x}} \right) $, is determined by optimizing the conditional PDF $ p_{\hat{\vec{x}} | \vec{y} }\left( \vecRvReal{\hat{x}} | \vecRvReal{y}  \right) $, statistically representing the mapping between the given data $ \vec{y} $ and the decompression result $ \hat{\vec{x}} $. Moreover, the rate is measured as the mutual information between $\hat{\vec{x}}$ and $ \vec{y} $, defined via 
\begin{IEEEeqnarray}{rCl}
	\label{eq:theoretic Gaussian analysis - mutual information definition}
	I\left( \vec{y}; \hat{\vec{x}} \right) = \int{ p_{\vec{y}, \hat{\vec{x}}}\left(  \vecRvReal{y}, \vecRvReal{\hat{x}} \right)  \log{ \frac{p_{\vec{y}, \hat{\vec{x}}} \left( \vecRvReal{y}, \vecRvReal{\hat{x}} \right)}{p_{\vec{y}}\left(  \vecRvReal{y}\right) p_{\hat{\vec{x}}}\left(  \vecRvReal{\hat{x}} \right) } } d \vecRvReal{y} d \vecRvReal{\hat{x}} } .
\end{IEEEeqnarray}
Then, the basic form of the complexity-regularized restoration optimization is expressed as 
\begin{problemform}[Basic Form]
	\label{problem: basic form}
	\begin{IEEEeqnarray}{rCl}
	\begin{aligned}
		& \underset{ p_{\hat{\vec{x}}|{\vec{y}}}  }{\text{min}}
		& & { I\left( \vec{y}; \hat{\vec{x}} \right) } \\
		& \text{subject to}
		& & E\left\lbrace \left\| { \vec{y}  - \mtx{H} \hat{\vec{x}} } \right\|_2^2  \right\rbrace = N \sigma_n^2 .
	\end{aligned}	
	\end{IEEEeqnarray}
\end{problemform}
Here the estimate rate is minimized while maintaining suitability to the degradation model (\ref{eq:theoretic Gaussian analysis - degradation model}) using a distortion constraint set to achieve an a-priori known total noise energy. In general, Problem \ref{problem: basic form} is complicated to solve since the distortion constraint considers  $ \hat{\vec{x}} $ through the degradation operator $ \mtx{H} $, while the rate is directly evaluated for $ \hat{\vec{x}} $.

The shift invariant operator $ \mtx{H} $ is a circulant $ N \times N $ matrix, thus, diagonalized by the $ N\times N $ Discrete Fourier Transform (DFT) matrix 
$ \mtx{F} $. The $ (k,l) $ component of the DFT matrix ($ k,l=0,...,N-1$) is  $\mtx{F}_{k,l}= W_N^{kl} $ where $ W_N \triangleq \frac{1}{\sqrt{N}} e^{-i2\pi/N} $.
Then, the diagonalization of $ \mtx{H} $ is expressed as 
\begin{IEEEeqnarray}{rCl}
	\label{eq:theoretic Gaussian analysis - diagonlization of H}
	\mtx{F} \mtx{H} \mtx{F}^* = \mtx{\Lambda}_H , 
\end{IEEEeqnarray}
where $ \mtx{\Lambda}_H $ is a diagonal matrix formed by the components $ h^F_k $ for $ k=0,...,N-1 $.
Using $ \mtx{\Lambda}_H $ we define the pseudoinverse of $ \mtx{H} $ as 
\begin{IEEEeqnarray}{rCl}
	\label{eq:theoretic Gaussian analysis - pseudoinverse of H}
	\mtx{H}^{+} =  \mtx{F}^* \mtx{\Lambda}_H^{+} \mtx{F} , 
\end{IEEEeqnarray}
where $ \mtx{\Lambda}_H^{+} $ is the pseudoinverse of $ \mtx{\Lambda}_H $, an $ N \times N $ diagonal matrix with the $ k^{th} $ diagonal element: 
\begin{IEEEeqnarray}{rCl}
	\label{eq:theoretic Gaussian analysis - pseudoinverse of Lambda_H - diagonal elements}
	h^{F,+}_k = \left\{ {\begin{array}{*{20}{c}}
			{\frac{1}{h^F_k}}&{,\text{for~~} h^F_k\ne 0 } \\ 
			0&{,\text{for~~} h^F_k = 0 . } 
		\end{array}} \right.  
\end{IEEEeqnarray} 
We denote by $N_{H}$ the number of nonzero diagonal elements in  $ \mtx{\Lambda}_H $, the rank of $ \mtx{H} $.

The first main result of our analysis states that Problem \ref{problem: basic form}, being the straightforward formulation for complexity-regularized restoration, is equivalent to the next problem. 
\begin{problemform}[Pseudoinverse-filtered input]
	\label{problem: pseudoinverse filtered input}
	\begin{IEEEeqnarray}{rCl}
		\begin{aligned}
			& \underset{ p_{\hat{\vec{x}}|{\tilde{\vec{y}}}}  }{\text{min}}
			& & { I\left( {\tilde{\vec{y}}}; \hat{\vec{x}} \right) } \\
			& \text{subject to}
			& & E\left\lbrace \left\| \mtx{H} \left(  { {\tilde{\vec{y}}}  -  \hat{\vec{x}} } \right) \right\|_2^2  \right\rbrace = N_{H} \sigma_n^2 , 
		\end{aligned}	
	\end{IEEEeqnarray}
\end{problemform}
where 
\begin{IEEEeqnarray}{rCl}
	\label{eq:theoretic Gaussian analysis - pseudoinverse filtered input definition}
	{\tilde{\vec{y}}} = \mtx{H}^{+} \vec{y}
\end{IEEEeqnarray}
is the pseudoinverse filtered version of the given degraded signal $ \vec{y} $.
One should note that Problem \ref{problem: pseudoinverse filtered input} has a more convenient form than Problem \ref{problem: basic form} since the distortion is an expected weighted squared error between the two random variables determining the rate. 
The equivalence of Problems \ref{problem: basic form} and \ref{problem: pseudoinverse filtered input} is proved in  Appendix \ref{appendix:Equivalence of Problems 1 and 2}.

In this section, $ \vec{x} $ is a cyclo-stationary Gaussian signal, hence, having a circulant autocorrelation matrix $ \mtx{R}_{\vec{x}} $. Consequently, and also because $ \mtx{H} $ is circulant, the deteriorated signal $ \vec{y} $ is also a cyclo-stationary Gaussian signal. Moreover, $ \mtx{H}^{+} $ is also a circulant matrix, thus, by (\ref{eq:theoretic Gaussian analysis - pseudoinverse filtered input definition}) the pseudoinverse filtering result, $ {\tilde{\vec{y}}} $, is also cyclo-stationary and zero-mean Gaussian. Specifically, the autocorrelation matrix of  $ {\tilde{\vec{y}}} $ is 
\begin{IEEEeqnarray}{rCl}
	\label{eq:theoretic Gaussian analysis - pseudoinverse filtered input - autocorrelation matrix}
	\mtx{R}_{\tilde{\vec{y}}} & = & \mtx{H}^{+} \mtx{R}_{\vec{y}} {\mtx{H}^{+}}^*
	\\ 
	& = &  \mtx{H}^{+} \mtx{H} \mtx{R}_{\vec{x}} {\mtx{H}}^* {\mtx{H}^{+}}^* + \sigma_n^2 \mtx{H}^{+} {\mtx{H}^{+}}^* ,  
\end{IEEEeqnarray}
and, as a circulant matrix, it is diagonalized by the DFT matrix yielding the eigenvalues 
\begin{IEEEeqnarray}{rCl}
	\label{eq:theoretic Gaussian analysis - pseudoinverse filtered input - autocorrelation eigenvalues}
	\lambda^{\left(\tilde{\vec{y}}\right)}_k = \left\{ {\begin{array}{*{20}{c}}
			{\lambda^{\left(\vec{x}\right)}_k  + \frac{\sigma_n^2}{\left| h^F_k \right| ^2}}&{,\text{for~~} h^F_k\ne 0 } \\ 
			0&{,\text{for~~} h^F_k = 0 . } 
		\end{array}} \right.  
\end{IEEEeqnarray}
The DFT-domain representation of $ {\tilde{\vec{y}}} $ is 
\begin{IEEEeqnarray}{rCl}
	\label{eq:theoretic Gaussian analysis - pseudoinverse filtered input - in DFT domain}
	{\tilde{\vec{y}}}^F = \mtx{F}{\tilde{\vec{y}}}, 
\end{IEEEeqnarray}
consisted of the coefficients $ \left\lbrace {\tilde{y}}^F_k \right\rbrace_{k=0}^{N-1} $, being independent zero-mean Gaussian variables with variances corresponding to the eigenvalues in (\ref{eq:theoretic Gaussian analysis - pseudoinverse filtered input - autocorrelation eigenvalues}).

Transforming Problem \ref{problem: pseudoinverse filtered input} to the DFT domain, where $ {\tilde{\vec{y}}} $ becomes a set of independent Gaussian variables to be coded under a joint distortion constraint, simplifies the optimization structure to the following separable form (see proof sketch in Appendix \ref{appendix:Equivalence of Problems 2 and 3}).
\begin{problemform}[Separable form in DFT domain]
	\label{problem: Separable Form in DFT domain}
	\begin{IEEEeqnarray}{rCl}
		\begin{aligned}
			& \underset{ \left\lbrace p_{\hat{x}^F_k|{\tilde{y}^F_k}} \right\rbrace_{k=0}^{N-1} }{\text{min}}
			& & { \sum\limits_{k=0 }^{N-1} I\left( {\tilde{y}^F_k}; \hat{x}^F_k \right) } \\
			& \text{subject to}
			& & \sum\limits_{k=0}^{N-1} {\left| h^F_k \right| ^2} E\left\lbrace  { \left| { {\tilde{y}^F_k}  -  \hat{x}^F_k } \right| ^2  } \right\rbrace  = N_{H} \sigma_n^2 , 
		\end{aligned}	
	\end{IEEEeqnarray}
\end{problemform}
where $ \left\lbrace {\hat{x}}^F_k \right\rbrace_{k=0}^{N-1} $ are the elements of $ {\hat{\vec{x}}}^F = \mtx{F}{\hat{\vec{x}}} $. Nicely, the separable distortion in Problem \ref{problem: Separable Form in DFT domain} considers each variable using a squared error that is weighted by the squared magnitude of the corresponding degradation-filter coefficient. 

The rate-distortion function of a single Gaussian variable with variance $ \sigma^2 $ has the known formulation \cite{cover2012elements}:
\begin{IEEEeqnarray}{rCl}
	\label{eq:theoretic Gaussian analysis - rate-distortion function of a Gaussian variable}
	R\left(D\right) = \left[ \frac{1}{2} \log \left( { \frac{\sigma^2}{D} } \right) \right]_+
\end{IEEEeqnarray}
evaluating the minimal rate for a squared-error allowed reaching up to $ D \ge 0 $. In addition, the operator $ \left[\cdot\right]_+ $ is defined for real scalars as $ \left[\alpha\right]_+ \triangleq \max\left\{\alpha, 0\right\} $, hence, $ R \left( D \right) = 0 $ for $ D \ge \sigma^2 $. 
Accordingly, the rate-distortion function of the Gaussian variable ${\tilde{y}^F_k}$ is 
\begin{IEEEeqnarray}{rCl}
	\label{eq:theoretic Gaussian analysis - rate-distortion function of DFT component}
	R_k\left(D_k\right) = \left[ \frac{1}{2} \log \left( { \frac{ \lambda^{\left(\tilde{\vec{y}}\right)}_k }{D_k} } \right) \right]_+
\end{IEEEeqnarray} 
where $ D_k $ denotes the maximal squared-error allowed for this component. 
Now, similar to the famous case of jointly coding independent Gaussian variables with respect to a regular (non-weighted) squared-error distortion \cite{cover2012elements}, we explicitly express Problem \ref{problem: Separable Form in DFT domain} as the following distortion-allocation optimization.
\begin{problemform}[Explicit distortion allocation]
	\label{problem: Gaussian distortion allocation}
\begin{IEEEeqnarray}{rCl}
	\label{eq:theoretic Gaussian analysis - Gaussian distortion allocation}
	\begin{aligned}
		& \underset{D_0,...,D_{N-1}}{\text{min}}
		& & \sum\limits_{k=0}^{N-1} { \left[ \frac{1}{2} \log \left( { \frac{ \lambda^{\left(\tilde{\vec{y}}\right)}_k }{D_k} } \right) \right]_+ } \\
		& \text{subject to}
		& & \sum\limits_{k=0}^{N-1} { \left| h^F_k \right|^2 D_k  } = N_{H} \sigma_n^2 
		\\
		& & & D_k \ge 0 ~~~,~k=0,...,N-1.
	\end{aligned}
\end{IEEEeqnarray}
\end{problemform}
The optimal distortion-allocation satisfying the last optimization is 
\begin{IEEEeqnarray}{rCl}
	\label{eq:theoretic Gaussian analysis - optimal distortion allocation}
	D_k^{opt} = \left\{ {\begin{array}{*{20}{c}}
			{\frac{ \sigma_n^2 }{\left| h^F_k \right|^2 }}&{,\text{for~~} h^F_k\ne 0 } \\ 
			0&{,\text{for~~} h^F_k = 0 } 
		\end{array}} \right.
\end{IEEEeqnarray}
and the associated optimal rates are 
\begin{IEEEeqnarray}{rCl}
	\label{eq:theoretic Gaussian analysis - optimal rate allocation}
	R_k^{opt} = \left\{ {\begin{array}{*{20}{c}}
			{\frac{1}{2} \log\left( \left| h^F_k \right|^2 \frac{\lambda^{(\vec{x})}_k  }{\sigma_n^2} + 1 \right)}&{,\text{for~~} h^F_k\ne 0 } \\ 
			0&{,\text{for~~} h^F_k = 0 . } 
		\end{array}} \right.  	
\end{IEEEeqnarray}
Results (\ref{eq:theoretic Gaussian analysis - optimal distortion allocation}) and (\ref{eq:theoretic Gaussian analysis - optimal rate allocation}) are proved in Appendix \ref{appendix:Solution of Problem 4}.

\subsection{Demonstration of The Explicit Results}
\label{subsec:Rate-Distortion Theoretic - Demonstration of The Explicit Results}

\begin{figure}[]
	\centering
	{\subfloat[]{\label{fig:signal_autocorrelation_matrix}\includegraphics[width=0.24\textwidth]{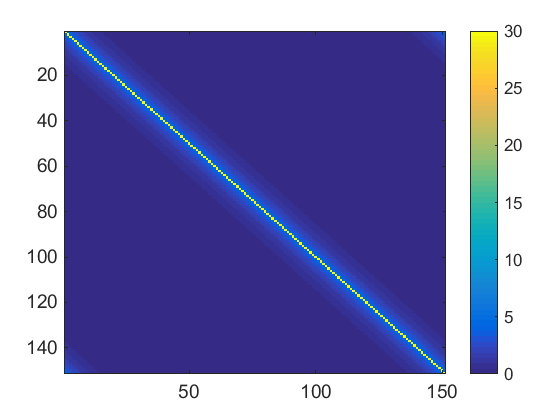}}}
	{\subfloat[]{\label{fig:signal_autocorrelation_matrix_DFT_decomposition}\includegraphics[width=0.24\textwidth]{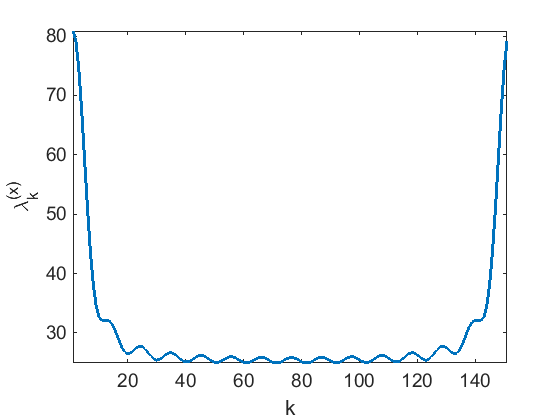}}}
	\caption{The autocorrelation of the cyclo-stationary Gaussian signal used in the demonstration. (a) The circulant autocorrelation matrix in the signal domain, and (b) the corresponding eigenvalues obtained using the DFT decomposition.} 
	\label{Fig:Gaussian signal for demonstration}
\end{figure}

\begin{figure*}[]
	\centering
	{\subfloat[]{\label{fig:gaussian_denoising__degradation_filter_signal_domain}\includegraphics[width=0.24\textwidth]{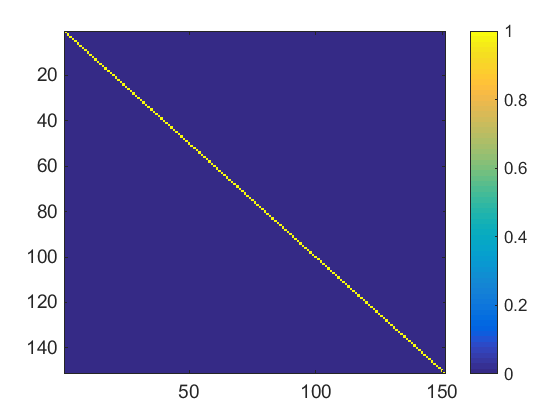}}}
	{\subfloat[]{\label{fig:gaussian_denoising__degradation_filter_DFT_coefficient_magnitudes}\includegraphics[width=0.24\textwidth]{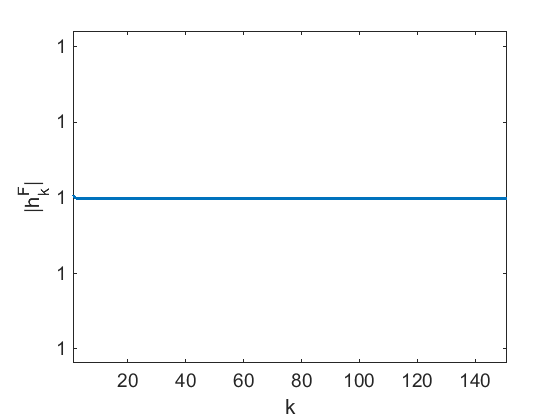}}}
	{\subfloat[]{\label{fig:gaussian_denoising__effective_waterfilling_demonstration}\includegraphics[width=0.24\textwidth]{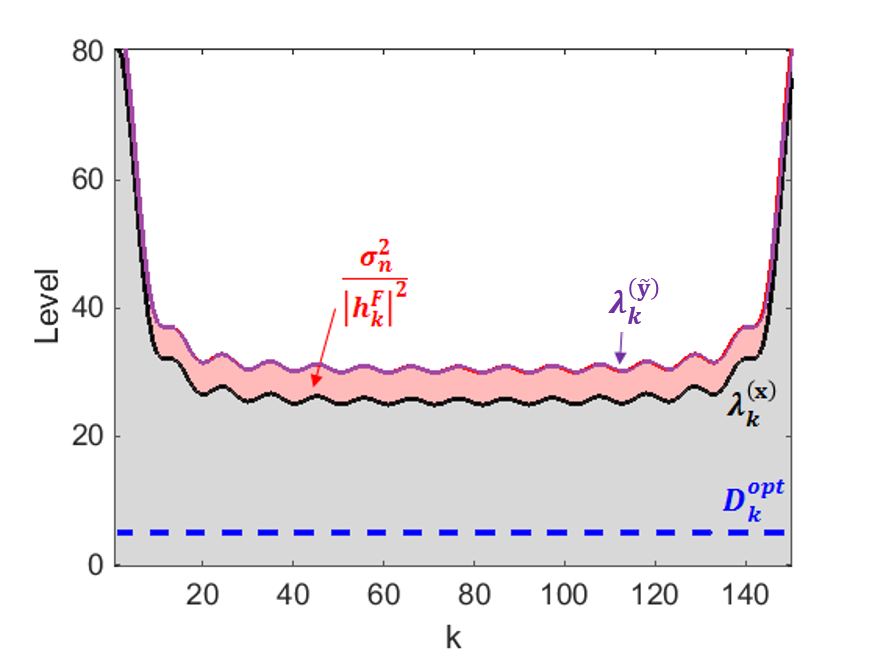}}}
	{\subfloat[]{\label{fig:gaussian_denoising__optimal_rate_allocation}\includegraphics[width=0.24\textwidth]{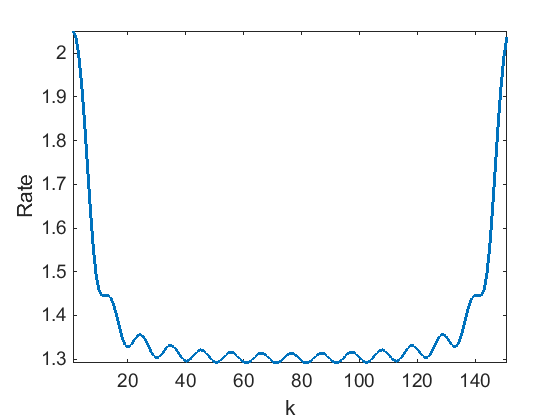}}}
	\caption{Demonstrating the theoretic results for a denoising problem with a noise level of $ \sigma_n^2 = 5 $. (a) The degenerated degradation operator in the signal domain $ \mtx{H} = \mtx{I} $. (b) DFT-domain magnitude of the degradation filter. (c) Optimal waterfilling solution in DFT domain. (d) Optimal rate allocation in DFT domain.} 
	\label{Fig:Gaussian signal - denoising demonstration}
\end{figure*}

Let us exemplify the optimal rate-distortion results (\ref{eq:theoretic Gaussian analysis - optimal distortion allocation})-(\ref{eq:theoretic Gaussian analysis - optimal rate allocation}) for a cyclo-stationary Gaussian signal, $ \vec{x} $, having the circulant autocorrelation matrix presented in Fig. \ref{fig:signal_autocorrelation_matrix}, corresponding to the eigenvalues $ \{ {\lambda}^{(\vec{x})}_k \} _{k=0}^{N-1} $ (Fig. \ref{fig:signal_autocorrelation_matrix_DFT_decomposition}) obtained by a DFT-based decomposition. 
We first examine the denoising problem, where the signal-domain degradation matrix is $ \mtx{H} = \mtx{I} $ (Fig. \ref{fig:gaussian_denoising__degradation_filter_signal_domain}) and its respective DFT-domain spectral representation consists of $ h^F_k = 1 $ for any $ k $ (see Fig. \ref{fig:gaussian_denoising__degradation_filter_DFT_coefficient_magnitudes}). The additive white Gaussian noise has a sample variance of $ \sigma_n^2 = 5 $.
Fig. \ref{fig:gaussian_denoising__effective_waterfilling_demonstration} exhibits the optimal distortion allocation using a reverse-waterfilling diagram, where the signal-energy distribution $ \{ {\lambda}^{(\vec{x})}_k \} _{k=0}^{N-1} $ (black solid line) and the additive noise energy (the light-red region) defining together the noisy-signal energy level (purple solid line) corresponding to $\lambda^{\left(\tilde{\vec{y}}\right)}_k = \lambda^{(\vec{x})}_k + \sigma_n^2$. The blue dashed line in Fig. \ref{fig:gaussian_denoising__effective_waterfilling_demonstration} shows the water level associated with the uniform distortion allocation. 
The optimal rate-allocation, corresponding to Fig. \ref{fig:gaussian_denoising__effective_waterfilling_demonstration} and Eq. (\ref{eq:theoretic Gaussian analysis - optimal rate allocation}), is presented in Fig. \ref{fig:gaussian_denoising__optimal_rate_allocation} showing that more bits are spent on components with higher signal-to-noise ratios.

\begin{figure*}[]
	\centering
	{\subfloat[]{\label{fig:gaussian_restoration__degradation_filter_signal_domain}\includegraphics[width=0.24\textwidth]{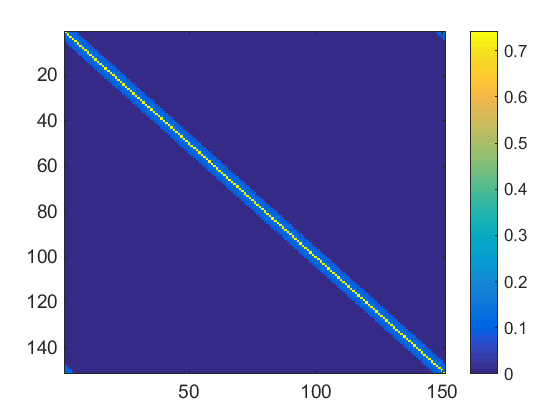}}}
	{\subfloat[]{\label{fig:gaussian_restoration__degradation_filter_DFT_coefficient_magnitudes}\includegraphics[width=0.24\textwidth]{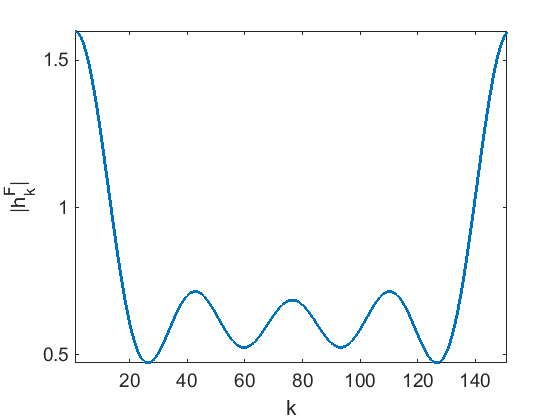}}}
	{\subfloat[]{\label{fig:gaussian_restoration__effective_waterfilling_demonstration}\includegraphics[width=0.24\textwidth]{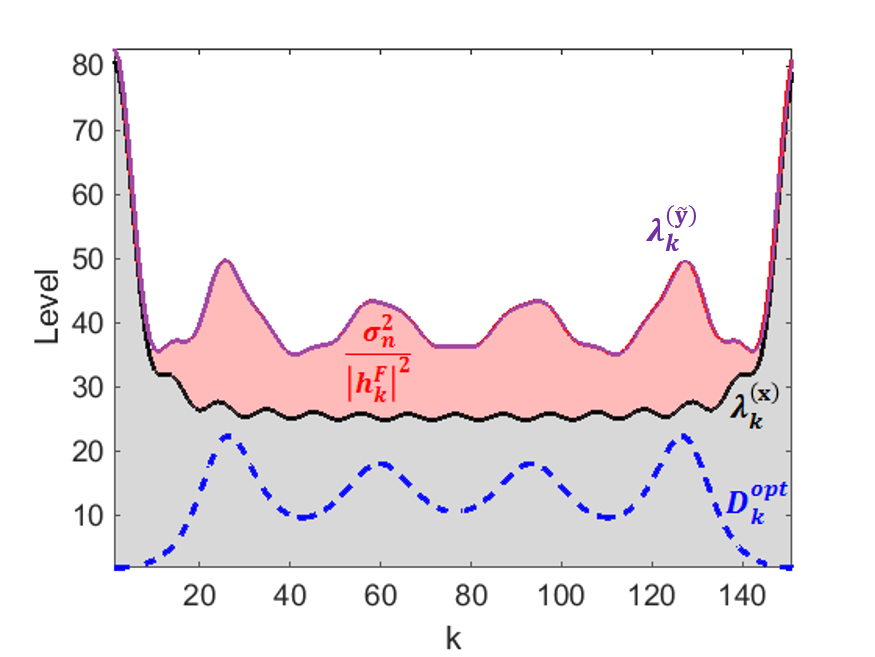}}}
	{\subfloat[]{\label{fig:gaussian_restoration__optimal_rate_allocation}\includegraphics[width=0.24\textwidth]{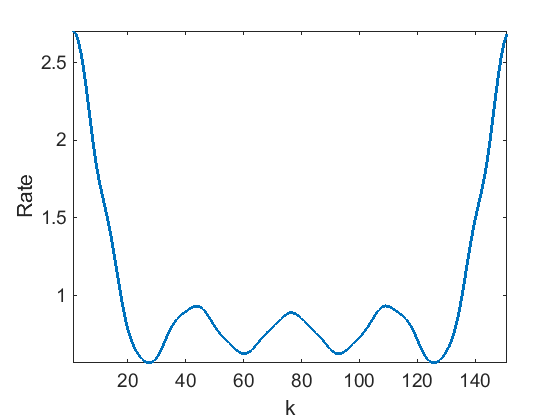}}}
	\caption{Demonstrating the theoretic results for a restoration problem with a noise level of $ \sigma_n^2 = 5 $. (a) The degradation operator in the signal domain $ \mtx{H} $. (b) DFT-domain magnitude of the degradation filter. (c) Optimal waterfilling solution in DFT domain. (d) Optimal rate allocation in DFT domain.} 
	\label{Fig:Gaussian signal - restoration demonstration}
\end{figure*}

Another example considers the same Gaussian signal described in Fig. \ref{Fig:Gaussian signal for demonstration} and the noise level of $ \sigma_n^2 = 5 $, but here the degradation operator is the circulant matrix shown in Fig. \ref{fig:gaussian_restoration__degradation_filter_signal_domain} having a DFT-domain representation given in magnitude-levels in Fig. \ref{fig:gaussian_restoration__degradation_filter_DFT_coefficient_magnitudes} exhibiting its frequency attenuation and amplification effects. 
The waterfilling diagram in Fig. \ref{fig:gaussian_restoration__effective_waterfilling_demonstration} includes the same level of signal energy (black solid line) as in the denoising experiment, but the effective additive noise levels and the allocated distortions are clearly modulated in an inversely proportional manner by the squared magnitude of the degradation operator. For instance, frequencies corresponding to degradation-filter magnitudes lower than 1 lead to increase in the effective noise-energy addition and in the allocated distortion. 
The optimal rate allocation (Fig. \ref{fig:gaussian_restoration__optimal_rate_allocation}) is affected by the signal-to-noise ratio and by the squared-magnitude of the degradation filter (see also Eq. (\ref{eq:theoretic Gaussian analysis - optimal rate allocation})), e.g., components that are attenuated by the degradation operator get less bits in the rate allocation.

\subsection{Conceptual Relation to The Proposed Approach}
\label{subsec:Rate-Distortion Theoretic - Conceptual Relation to The Proposed Approach}

As explained at the beginning of this section, theoretical and practical solutions may include different implementations of the same general ideas. Accordingly, connections between theory and practice should be established by pointing on high-level analogies.
Our rate-distortion analysis (for a Gaussian signal and a LSI degradation operator) showed that the optimal complexity-regularized restoration relies on two prominent ideas: pseudoinverse filtering of the degraded input, and compression with respect to a squared-error metric that is weighted based on the degradation-filter squared-magnitude (considering the DFT-domain procedure). 
We will now turn to explain how these two concepts connect to more general themes having different realizations in the practical approach proposed in Section \ref{sec:Proposed Methods} \footnote{Since the differences between Algorithm \ref{Algorithm:Proposed Method Non-overlapping} and Algorithms \ref{Algorithm:Proposed Method Overlapping Blocks} and \ref{Algorithm:Proposed Method Overlapping Blocks with Robust Dual Variables} are for a shift-invariance purpose, an issue that we do not concern in this section, we compare our theoretic results only to Algorithm \ref{Algorithm:Proposed Method Non-overlapping}.}. 

$\bullet$ \textit{\textbf{Design Concept \#1:} Apply simple restoration filtering}.
The general idea of using an elementary restoration filter is implemented in the Gaussian case as pseudoinverse filtering. 
Correspondingly, our practical approach relies on a simple filtering mechanism, extending the pseudoinverse filter as explained next. Stage 6 of Algorithm \ref{Algorithm:Proposed Method Non-overlapping} is an $ \ell_2 $-constrained deconvolution filtering that its analytic solution can be rewritten, using the relation $ \mtx{H}^*\left( \mtx{I} - \mtx{H}\mtx{H^*} \right) \nolinebreak=\nolinebreak \mtx{0} $, as (see proof in Appendix \ref{appendix:Equivalent Forms of Stage 4 of Algorithm 1})
\begin{IEEEeqnarray}{rCl}
	\label{eq:theory-practice relation - interpreting first stage}
	\hat{ \vec{x}}^{(t)} = \left(  \mtx{H}^* \mtx{H} + \frac{\beta}{2} \mtx{I}  \right)^{-1} \left( \mtx{H}^* \mtx{H} \tilde{\vec{y}} + \frac{\beta}{2} \tilde{\vec{z}}^{(t)}   \right) .~
\end{IEEEeqnarray}
As before, $ \tilde{\vec{y}} = \mtx{H}^+ \vec{y} $, i.e., the pseudoinverse-filtered version of $ \vec{y} $. The expression (\ref{eq:theory-practice relation - interpreting first stage}) can be interpreted as an initial pseudoinverse filtering of the degraded input, followed by a simple weighted averaging with $\tilde{\vec{z}}^{(t)}$ (that includes the decompressed signal obtained in the last iteration).
Evidently, the filtering in (\ref{eq:theory-practice relation - interpreting first stage}) is determined by the $\beta$ value, specifically, for $ \beta=0 $ the estimate coincides with the pseudoinverse filtering solution and for a larger $\beta$ it is closer to $\tilde{\vec{z}}^{(t)}$.

$\bullet$ \textit{\textbf{Design Concept \#2:} Compress by promoting higher quality for signal-components matching to higher $ h $-operator magnitudes}.  
This principle is realized in the theoretic Gaussian case as weights attached to the squared-errors of DFT-domain components (see Problems \ref{problem: Separable Form in DFT domain} and \ref{problem: Gaussian distortion allocation}). Since the weights, ($ \left\lvert h_0^F \right\rvert ^2 , ..., \left\lvert h_{N-1}^F \right\rvert ^2 $), are the squared magnitudes of the corresponding degradation-filter coefficients, in the compression of the pseudoinverse-filtered input the distortion is spread unevenly being larger where the degradation filter-magnitude is lower. 
Remarkably, this concept is implemented differently in the proposed procedure (Algorithm \ref{Algorithm:Proposed Method Non-overlapping}) where regular compression techniques, optimized for the squared-error distortion measure, are applied on the filtering result of the preceding stage. We will consider the essence of the effective compression corresponding to these two stages together. Let us revisit (\ref{eq:theory-practice relation - interpreting first stage}), expressing stage 6 of Algorithm \ref{Algorithm:Proposed Method Non-overlapping}. Assuming $ \mtx{H} $ is a circulant matrix, we can transform (\ref{eq:theory-practice relation - interpreting first stage}) into its Fourier domain representation 
\begin{IEEEeqnarray}{rCl}
	\label{eq:theory-practice relation - interpreting first stage - circulant H}
	\hat{ \vec{x}}^{F,(t)} = \left(  \mtx{\Lambda}_H^* \mtx{\Lambda}_H + \frac{\beta}{2} \mtx{I}  \right)^{-1} \left( \mtx{\Lambda}_H^* \mtx{\Lambda}_H \tilde{\vec{y}}^{F} + \frac{\beta}{2} \tilde{\vec{z}}^{F,(t)}   \right) \nonumber\\
\end{IEEEeqnarray}
where $ \hat{ \vec{x}}^{F,(t)} $ and $ \tilde{\vec{z}}^{F,(t)} $ are the Fourier representations of $ \hat{ \vec{x}}^{(t)} $ and $ \tilde{\vec{z}}^{(t)} $, respectively. Furthermore, (\ref{eq:theory-practice relation - interpreting first stage - circulant H}) reduces to the componentwise formulation 
\begin{IEEEeqnarray}{rCl}
	\label{eq:theory-practice relation - interpreting first stage - circulant H - component}
	\hat{ {x}}^{F,(t)}_k = \frac{\left\lvert h_k^F \right\rvert ^2  \tilde{y}^{F}_k + \frac{\beta}{2} \tilde{z}^{F,(t)}_k }{ \left\lvert h_k^F \right\rvert ^2  + \frac{\beta}{2} }
\end{IEEEeqnarray}
where  $ \hat{ {x}}^{F,(t)}_k $ and $ \tilde{z}^{F,(t)}_k $ are the $ k^{th} $ Fourier coefficients of $ \hat{ \vec{x}}^{F,(t)} $ and $ \tilde{\vec{z}}^{F,(t)} $, respectively.
Equation (\ref{eq:theory-practice relation - interpreting first stage - circulant H - component}) shows that signal elements (of the pseudoinverse-filtered input) corresponding to degradation-filter components of weaker energies will be retracted more closely to the respective components of $ \tilde{z}^{F,(t)}_k $ -- thus, will be farther from $\tilde{y}^{F}_k$, yielding that the corresponding components in the standard compression applied in the next stage of this iteration will be of a relatively lower quality with respect their matching components of $\tilde{y}^{F}_k$ (as they were already retracted relatively far from them in the preceding deconvolution stage).

To conclude this section, we showed that the main architectural ideas expressed in theory (for the Gaussian case) appear also in our practical procedure. The iterative nature of our methods (Algorithms \ref{Algorithm:Proposed Method Non-overlapping}-\ref{Algorithm:Proposed Method Overlapping Blocks with Robust Dual Variables}) as well as the desired shift-invariance property provided by Algorithms \ref{Algorithm:Proposed Method Overlapping Blocks}-\ref{Algorithm:Proposed Method Overlapping Blocks with Robust Dual Variables} are outcomes of treating real-world scenarios such as non-Gaussian signals, general linear degradation operators, and computational limitations leading to block-based treatments -- these all relate to practical aspects, hence, do not affect the fundamental treatment given in this section.

\section{Experimental Results}
\label{sec:Experimental Results}

In this section we present experimental results for image restoration. Our main study cases include deblurring and inpainting using the image-compression profile of the HEVC standard (in its BPG implementation \cite{hevc_software_bpg}). We also provide evaluation of our method in conjunction with the JPEG2000 technique for the task of image deblurring.

We empirically found it sufficient to consider only a part of all the shifts, i.e., a portion of $ \mathcal{B}^* $. When using HEVC, the limited amount of shifts is compensated by the compression architecture that employs inter-block spatial predictions, thus, improves upon methods relying on independent block treatment.
The shifts are defined by the rectangular images having their upper-left corner pixel relatively close to the upper-left corner of the full image, and their bottom-right corner pixel coincides with that of the full image. This extends the mathematical developments in Section \ref{sec:Proposed Methods} as practical compression handles arbitrarily sized rectangular images.

Many image regularizers have visual interpretation, for example, the classical image-smoothness evaluation. In our framework, the regularization part in (\ref{eq:complexity-regularized restoration - full signal}) measures the complexity in terms of the compression bit-cost with respect to a specific compression architecture, designed based on some image model. 
Our complexity regularization also has a general visual meaning since, commonly, low bit-cost compressed images tend to be overly smooth or piecewise-smooth.

\subsection{Image Deblurring}
\label{subsec:Deblurring}

Here we consider two deterioration settings taken from \cite{danielyan2012bm3d}. The first setting, denoted here also as `Set. 1', considers a noise variance $ \sigma_n^2 = 2 $ coupled with a blur operator defined by the two-dimensional point-spread-function (PSF) $ h(x_1,x_2)=1/{(1+x_1^2+x_2^2)} $ for $ x_1,x_2=-7,...,7 $, and zero-valued otherwise. 
The second setting, denoted here also as `Set. 2' (named in \cite{danielyan2012bm3d} as `Scenario 3'), considers a noise variance $ \sigma_n^2 \approx 0.3 $ joint with a blur operator defined by the two-dimensional uniform blur PSF of size $ 9\times 9 $. 

\begin{figure}[]
	\centering
	{\subfloat[]{\label{fig:empirical_analysis_cost_evolution_deblurring_shift1_cameraman}\includegraphics[width=0.15\textwidth]{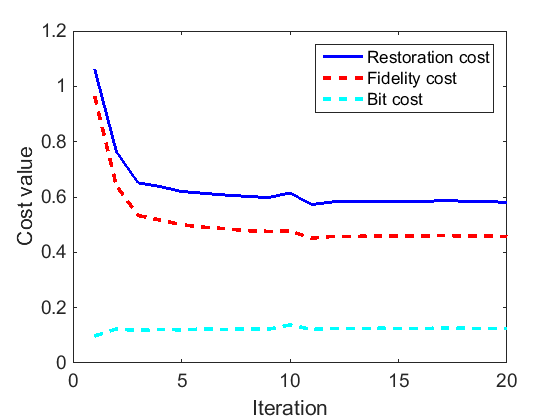}}}
	{\subfloat[]{\label{fig:empirical_analysis_psnr_evolution_deblurring_shift1_cameraman}\includegraphics[width=0.15\textwidth]{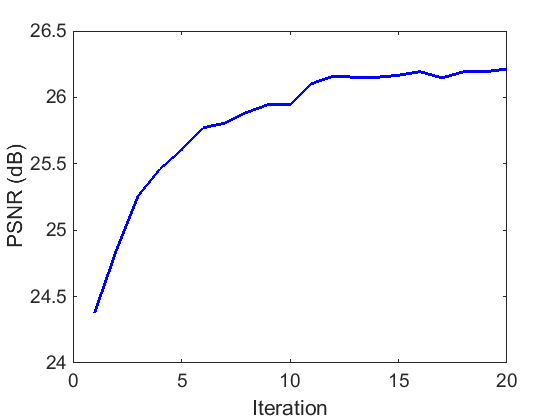}}}
	{\subfloat[]{\label{fig:empirical_analysis_compression_factor_evolution_deblurring_shift1_cameraman}\includegraphics[width=0.15\textwidth]{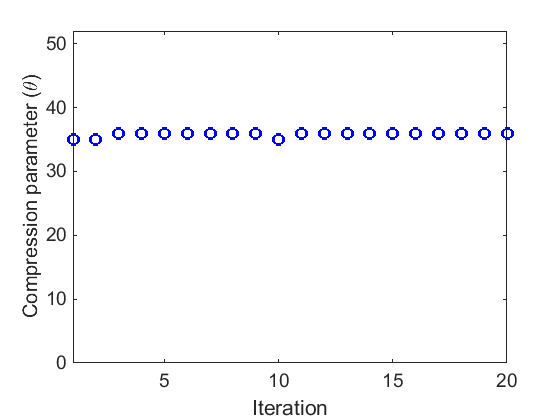}}}
	\caption{Empirical analysis for deblurring the Cameraman image using Algorithm \ref{Algorithm:Proposed Method Non-overlapping} (i.e., without overlapping blocks and shifted grids). The parameter settings here are: $ \mu = \left( 6.67\times 10^{-6} \right) $ and $ \beta = 0.01 $.} 
	\label{Fig:empirical_analysis_for_deblurring_shift1_cameraman}
\end{figure}

\begin{figure}[]
	\centering
	{\subfloat[]{\label{fig:empirical_analysis_cost_evolution_deblurring_shift3_cameraman}\includegraphics[width=0.15\textwidth]{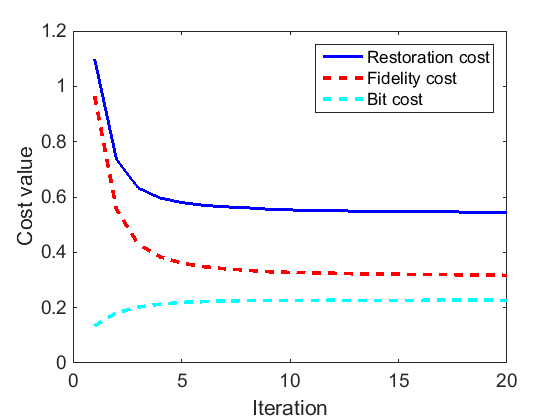}}}
	{\subfloat[]{\label{fig:empirical_analysis_psnr_evolution_deblurring_shift3_cameraman}\includegraphics[width=0.15\textwidth]{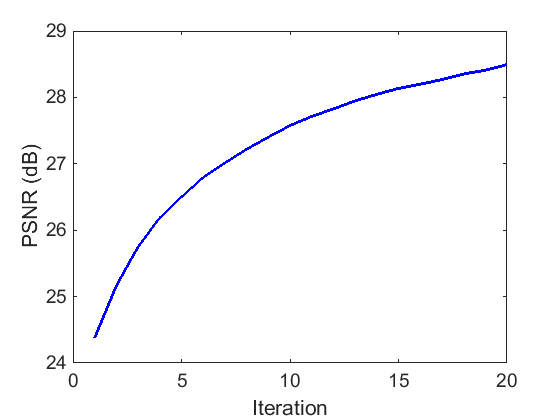}}}	
	{\subfloat[]{\label{fig:empirical_analysis_compression_factor_evolution_deblurring_shift3_cameraman}\includegraphics[width=0.15\textwidth]{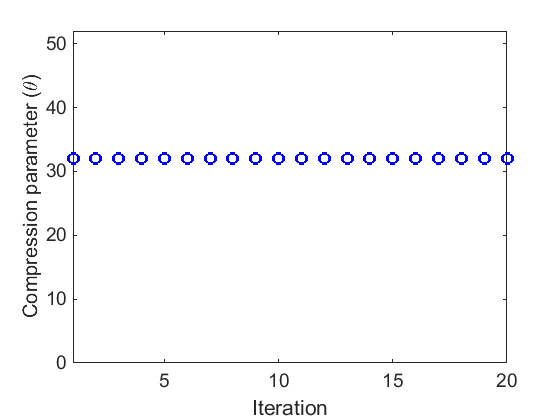}}}
	\caption{Empirical analysis for deblurring the Cameraman image using Algorithm \ref{Algorithm:Proposed Method Overlapping Blocks} considering 9 shifted block grids. The parameter settings here are: $ \mu = \frac{ \left( 6.67\times 10^{-6} \right) }{ \text{number of shifts} } $ and $ \beta = 0.01 $.} 
	\label{Fig:empirical_analysis_for_deblurring_shift3_cameraman}
\end{figure}

We precede the deblurring experiments with empirical evaluations of four important aspects of the proposed method.
\subsubsection{Iterative Reduction of the Fundamental Restoration Cost}
In Section \ref{sec:Proposed Methods} we established the basic optimization problems for restoration by regularizing the bit-costs of the non-overlapping and the overlapping blocks of the estimate (see (\ref{eq:complexity-regularized restoration - expressing blocks}) and (\ref{eq:complexity-regularized restoration - overlapping blocks}), respectively). As explained above, these two fundamental optimization tasks cannot be directly addressed and, therefore, we developed the ADMM-based Algorithms \ref{Algorithm:Proposed Method Non-overlapping}-\ref{Algorithm:Proposed Method Overlapping Blocks with Robust Dual Variables}, that iteratively employ simpler optimization problems. Figures \ref{fig:empirical_analysis_cost_evolution_deblurring_shift1_cameraman} and \ref{fig:empirical_analysis_cost_evolution_deblurring_shift3_cameraman} demonstrate that, for appropriate parameter settings, the fundamental optimization cost reduces in each iteration. The provided figures also show the fidelity term, $ \left\| {  \mtx{H} \vec{x} - \vec{y} } \right\|_2^2 $, and the regularizing bit-cost (multiplied by $ \mu $) of each iteration.

\subsubsection{Iterative Improvement of the Restored Image}
The fundamental optimization costs in (\ref{eq:complexity-regularized restoration - expressing blocks}) and (\ref{eq:complexity-regularized restoration - overlapping blocks}) include the fidelity term $ \left\| {  \mtx{H} \vec{x} - \vec{y} } \right\|_2^2 $ that considers the candidate estimate $\vec{x}$ and the given degraded signal $ \vec{y} $. However, the ultimate goal of the restoration process is to produce an estimate $\vec{x}$ that will be close to the original (unknown!) signal $\vec{x}_0$. It is common to evaluate proposed methods in experiment settings where $\vec{x}_0$ is known and used only for the evaluation of the squared error $ \left\| {  \vec{x} - \vec{x}_0 } \right\|_2^2 $ or its corresponding PSNR.
Accordingly, it is a desired property that our iterative methods will provide increment in the PSNR along the iterations and, indeed, Figures \ref{fig:empirical_analysis_psnr_evolution_deblurring_shift1_cameraman} and \ref{fig:empirical_analysis_psnr_evolution_deblurring_shift3_cameraman} show that this is achievable for appropriate parameter settings (the use of improper parameters may lead to unwanted decrease of the PSNR starting at some unknown iteration that, however, can be detected in many cases by heuristic divergence rules based on the dual variables used in the ADMM process). Interestingly, for some parameter settings, the PSNR may increase with the iterations, whereas the fundamental restoration cost will not necessarily consistently decrease. The last behavior may result from the fact the our optimization problem (with respect to a standard compression technique) is discrete, non-linear, and usually not convex and, therefore, the convergence guarantees of the ADMM \cite{boyd2011distributed} do not hold here for the fundamental restoration cost.

\subsubsection{The Optimal Compression Parameter}
\begin{figure}[]
	\centering
	{\subfloat[]{	\label{fig:empirical_analysis_deblurring_ADMM_JPEG2000_compression_ratios_cameraman}{\includegraphics[width=0.23\textwidth]{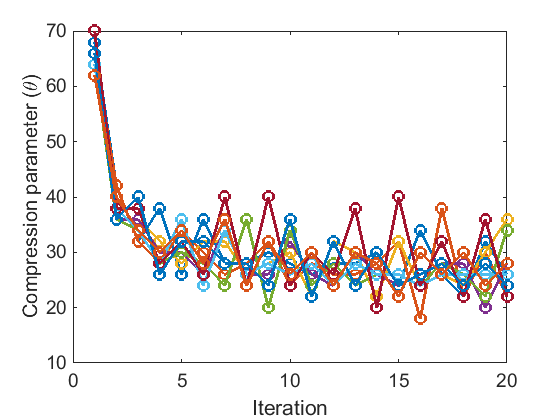}}}}
	\caption{Empirical analysis for deblurring the Cameraman image using Algorithm \ref{Algorithm:Proposed Method Overlapping Blocks} with JPEG2000 considering 9 shifted block grids. The compression parameter is the compression ratio given to the JPEG2000 compression. The parameter settings here are: 9 shifts, $ \mu = \frac{ \left( 1.33\times 10^{-4} \right)}{\text{number of shifts} } $, $ \beta = 25\times 10^{-4} $. } 
\end{figure}

Another question of practical importance is the value of the parameter $ \theta $, determining the compression level of the standard technique utilized in the proposed Algorithms \ref{Algorithm:Proposed Method Non-overlapping}-\ref{Algorithm:Proposed Method Overlapping Blocks with Robust Dual Variables}. Recall that the ADMM-based developments in Section \ref{sec:Proposed Methods} led to an iterative procedure including a stage of Lagrangian rate-distortion optimization operated for a Lagrange multiplier $ \lambda \triangleq \frac{2\mu}{\beta} $ and, then, we replaced this optimization with application of a standard compression-decompression technique operated based on a parameter $ \theta $. It is clear that $ \theta $ is a function of $ \lambda $. In the particular case where the standard compression has the Lagrangian form from our developments, then $ \theta = \lambda $, however, this is not the general case. 
For an arbitrary compression technique, we assume that its parameter $ \theta $ has $ K $ possible values $ \theta_1,...,\theta_K $ (for example, the HEVC standard supports 52 values for its quantization parameter), then, for a given $ \lambda \triangleq \frac{2\mu}{\beta} $ the required $ \theta $ value in stage 8 of Algorithm \ref{Algorithm:Proposed Method Non-overlapping} can be determined via 
\begin{IEEEeqnarray}{rCl}
	\label{eq:experiments - general setting of theta - rate-distortion optimization}
\theta^{(t)}_{\lambda,opt} = \mathop {{\text{argmin}}}\limits_{\theta \in \left\lbrace \theta_1,...,\theta_K \right\rbrace  }  { \left\| {  \tilde{\vec{x}}^{(t)} - \hat{\vec{z}}^{(t)}_{\theta} }\right\|_2^2} + \lambda r_{tot,\theta} .
\end{IEEEeqnarray}
where 
$  \hat{\vec{z}}^{(t)}_{\theta } = {CompressDecompress}_{\theta}\left( \tilde{\vec{x}}^{(t)} \right) $
is the decompressed signal and $ r_{tot,\theta} $ is the associated compression bit-cost.
We present here experiments (see Figs. \ref{Fig:empirical_analysis_for_deblurring_shift1_cameraman} and \ref{Fig:empirical_analysis_for_deblurring_shift3_cameraman}) for Algorithm \ref{Algorithm:Proposed Method Non-overlapping} and \ref{Algorithm:Proposed Method Overlapping Blocks} that in each iteration optimize the $ \theta $ value corresponding to $ \lambda \triangleq \frac{2\mu}{\beta} $ based on procedures similar to (\ref{eq:experiments - general setting of theta - rate-distortion optimization}).
Nicely, it is shown in Figures \ref{fig:empirical_analysis_compression_factor_evolution_deblurring_shift1_cameraman} and \ref{fig:empirical_analysis_compression_factor_evolution_deblurring_shift3_cameraman} that, for the HEVC compression used here, the best $ \theta $ values along the iterations are nearly the same (for a specific restoration task). This important property may be a result of the fact that HEVC extensively relies on Lagrangian rate-distortion optimizations (although in much more complex forms than those presented in Section \ref{sec:Proposed Methods}). Accordingly, in order to reduce the computational load, in the experiments shown below we will use a constant compression parameter given as an input to our methods. Interestingly, when examining the optimal compression parameters (compression ratios in this case) for the JPEG2000 method that applies wavelet-based transform coding, there is a decrease in the optimal compression ratio along the iterations (see Fig. \ref{fig:empirical_analysis_deblurring_ADMM_JPEG2000_compression_ratios_cameraman}). Accordingly, in order to reduce the computational load in the experiments below, we employed a predefined rule for reducing the JPEG2000 compression ratio along the iterations.
Importantly, when we use the sub-optimal predefined rules for setting the compression parameter $ \theta $ values (see Table \ref{table:Experimental Results - Parameter Setting} for the settings used in our evaluation comparisons in Table \ref{table:Deblurring - PSNR Comparison}) we do not longer need to set a value for $ \mu $ (since, in these cases, $ \mu $ is practically unused).

\begin{table} []
	\caption{Parameter Settings Used for the Deblurring and Inpainting Results in Tables \ref{table:Deblurring - PSNR Comparison} and \ref{table:Experimental Results - Inpainting}}
	\renewcommand{\arraystretch}{1.1}
	\label{table:Experimental Results - Parameter Setting}
	\centering
	\resizebox{\columnwidth}{!}{%
		\begin{tabular}{|c||c|c|c|c|}
			\hline
			& \shortstack{\\Maximal\\Number of Iterations} & \shortstack{\\Number of\\Shifts} & $ \beta $ & \shortstack{\\Compression Parameter\\$ \theta $} \\
			\hline\hline		                                       
			
			\shortstack{Deblurring\\Algorithm 2 and 3\\with HEVC}  & \shortstack{10 for `Set. 1'\\ 15 for `Set. 2' } & 400 & \shortstack{$ 5\times 10^{-3} $ for `Set. 1'\\ $ 10^{-3} $ for `Set. 2' } & Fixed on 40 \\
			\hline		                                               		                       
			\shortstack{Deblurring\\Algorithm 2 and 3\\with JPEG2000}  & 10  & 3600 & \shortstack{$ 5\times 10^{-3} $ for `Set. 1'\\ $ 25\times 10^{-4} $ for `Set. 2' } & \shortstack{Start at 90, decrease\\in 10 in each iteration\\keep fix when arriving to 10 }   \\
			\hline		                                               		                       
			\hline		                                               		                       
			\shortstack{Inpainting\\Algorithm 2 and 3\\with HEVC}  & 35 & 400 & $\frac{0.1}{ \text{number of shifts}  }$  & \shortstack{Fixed on 35 for Algorithm 2\\ Fixed on 40 for Algorithm 3 } \\
			\hline		                                               		                       
			\shortstack{Inpainting\\Algorithm 2\\with JPEG2000}  &  \shortstack{10 for Algorithm 2\\ 20 for Algorithm 3 } & 3600 & \shortstack{0 for Algorithm 2\\ $\frac{0.2}{ \text{number of shifts} }$ for Algorithm 3\\~ }& \shortstack{\\For Algorithm 2:\\Start at 78, decrease\\in 2 in each iteration\\keep fix when arriving to 25\\\\For Algorithm 3:\\Start at 98, decrease\\in 2 in each iteration\\keep fix when arriving to 50 } \\
			\hline		                                               		                       
			
		\end{tabular}
	}
\end{table}

\begin{table*} []
	\renewcommand{\arraystretch}{1.3}
	\caption{Deblurring: PSNR Comparison (The Three Best Results in Each Column Appear in Bold Text)}
	\label{table:Deblurring - PSNR Comparison}
	\centering
	\begin{tabular}{|c||c|c|c|c|c|c|c|c|}
		\hline
		\multirow{2}{*}{\bfseries } & \multicolumn{2}{|c|}{\bfseries \shortstack{\\Cameraman\\256x256}} &  \multicolumn{2}{|c|}{\bfseries\shortstack{\\House\\256x256}} & \multicolumn{2}{|c|}{\bfseries\shortstack{\\Lena\\512x512}} & \multicolumn{2}{|c|}{\bfseries\shortstack{\\Barbara\\512x512}} \\
		\cline{2-9}
		&  Set. 1 & Set. 2 & Set. 1 & Set. 2 & Set. 1 & Set. 2 & Set. 1 & Set. 2 \\
		\hline\hline
		Input PSNR & 22.23 & 20.76& 25.61 & 24.11 & 27.25 & 25.84 & 23.34 & 22.49 \\
		\hline\hline
		ForWaRD \cite{neelamani2004forward} & 28.99 & 28.10 & 32.96 & 33.67 & 33.30 & 32.81 & 27.03 & 26.51 \\
		\hline
		SV-GSM \cite{guerrero2008image} & 29.68 & 28.09 & 34.25 & 33.15 & - & - & 30.19 & 27.56 \\
		\hline
		BM3DDEB \cite{dabov2008image} & \textbf{30.42} & 29.10 & \textbf{34.93} & 34.96 & \textbf{35.20} & 33.81 & \textbf{31.14} & \textbf{28.35} \\
		\hline
		TVMM \cite{oliveira2009adaptive} & 29.64 & 29.30 & 33.59 & 34.50 & 33.61 & 33.31 & 26.44 & 25.98 \\		
		\hline
		CGMK \cite{chantas2010variational}	& 30.03 & 29.91 & 33.92 & 34.86 & 34.01 & 33.70 & 25.79 & 26.04 \\
		\hline
		IDD-BM3D \cite{danielyan2012bm3d}	& \textbf{31.08} & \textbf{31.21} & \textbf{35.56} & \textbf{37.00} & \textbf{35.22} & \textbf{34.75} & \textbf{30.98} & \textbf{28.54} \\
		\hline
		EPLL \cite{zoran2011learning}	& 29.40 & 29.54 & 33.88 & 35.87 & \textbf{34.64} & 34.39 & 27.62 & 26.78 \\
		\hline
		\shortstack{\\Proposed Algorithm 2\\with JPEG2000}	& 28.99 & 27.48 & 33.55 & 33.89 & 34.11 & 32.56 & 26.42 & 25.14 \\
		\hline
		\shortstack{\\Proposed Algorithm 3\\with JPEG2000}	& 29.28 & 28.10 & 33.07 & 34.01 & 34.38 & 32.60 & 26.31 & 25.18 \\
		\hline	
		\shortstack{\\Proposed Algorithm 2\\with HEVC}	& 30.19 & \textbf{30.20} & 33.17 & \textbf{36.47} & 33.32 & \textbf{34.40} & 29.83 & \textbf{27.74} \\
		\hline	
		\shortstack{\\Proposed Algorithm 3\\with HEVC}  & \textbf{30.35} & \textbf{30.14} & \textbf{34.37} & \textbf{36.57} & 34.55 & \textbf{34.41} & \textbf{30.20} & 27.72 \\
		\hline	
	\end{tabular}
\end{table*}

\subsubsection{Restoration Improvement for Increased Number of Image Shifts}
In our experiments we noticed that the restoration quality improves as more shifts are used, however, at some point the added gain due to the added shifts becomes marginal. As an example to the benefits due to shifts see Figs. \ref{fig:empirical_analysis_psnr_evolution_deblurring_shift1_cameraman} and \ref{fig:empirical_analysis_psnr_evolution_deblurring_shift3_cameraman}, where the PSNR obtained for deblurring the Cameraman image using Algorithm 2 and 9 shifts is about 2dB higher than the PSNR obtained using Algorithm 1 (i.e., without additional shifts).

\begin{figure*}[]
	\centering
	{\subfloat[{Original}]{\label{fig:deblurring - cameraman - true image}\includegraphics[width=0.24\textwidth]{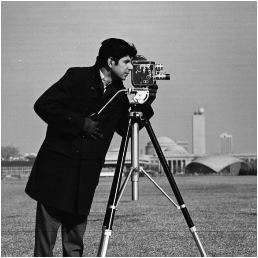}}}
	{\subfloat[{Deteriorated}]{\label{fig:deblurring - cameraman - blurry image}\includegraphics[width=0.24\textwidth]{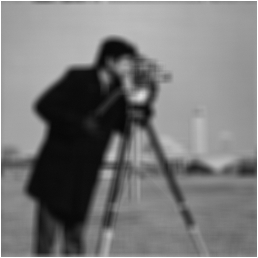}}}
	{\subfloat[{Proposed Restoration - JPEG2000}]{\label{fig:deblurring - cameraman - proposed restoration - JPEG2000}\includegraphics[width=0.24\textwidth]{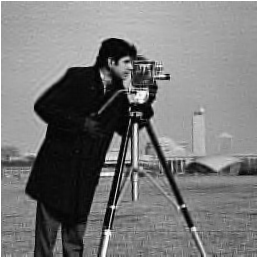}}}
	{\subfloat[{Proposed Restoration - HEVC}]{\label{fig:deblurring - cameraman - proposed restoration}\includegraphics[width=0.24\textwidth]{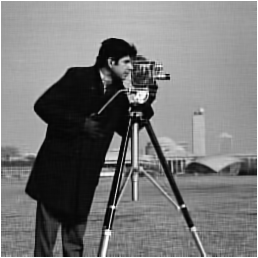}}}
	\caption{The deblurring experiment (settings \#2) for the Cameraman image ($ 256\times 256 $). (a) The underlying image. (b) Degraded image (20.76 dB). ~~(c) Restored image using Algorithm 3 with JPEG2000 compression (28.10 dB). (d) Restored image using Algorithm 3 with HEVC compression (30.14 dB). } 
	\label{Fig:Deblurring results - cameraman}
\end{figure*}
\begin{figure*}[]
	\centering
	{\subfloat[{Original}]{\label{fig:deblurring - barbara - true image}\includegraphics[width=0.24\textwidth]{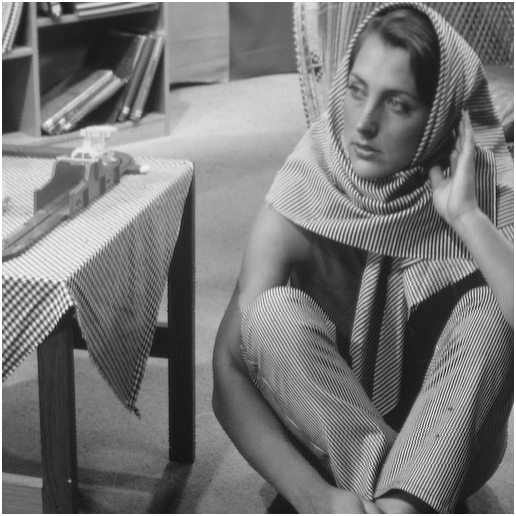}}}
	{\subfloat[{Deteriorated}]{\label{fig:deblurring - barbara - blurry image}\includegraphics[width=0.24\textwidth]{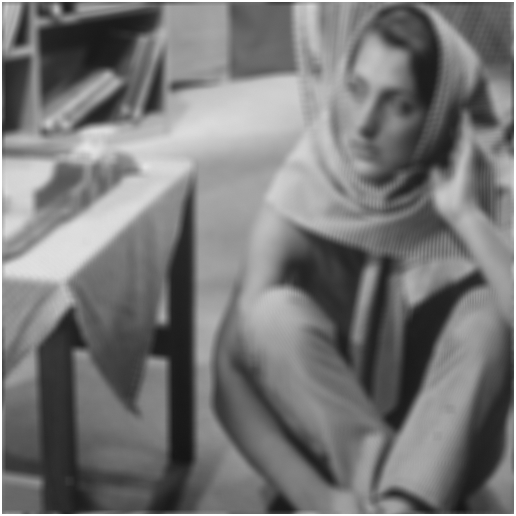}}}
	{\subfloat[{Proposed Restoration - JPEG2000}]{\label{fig:deblurring - barbara - proposed restoration - JPWG2000}\includegraphics[width=0.24\textwidth]{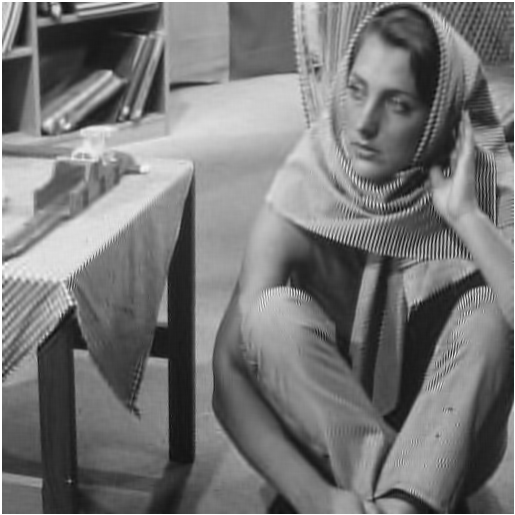}}}
	{\subfloat[{Proposed Restoration - HEVC}]{\label{fig:deblurring - barbara - proposed restoration}\includegraphics[width=0.24\textwidth]{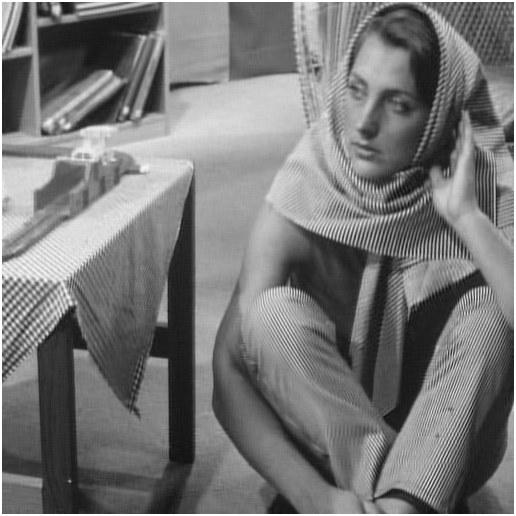}}}
	\caption{The deblurring experiment (settings \#2) for the Barbara image ($ 512\times 512 $). (a) The underlying image. (b) Degraded image (22.49 dB). (c) Restored image using Algorithm 3 with JPEG2000 compression (25.18 dB). (d) Restored image using Algorithm 3 with HEVC compression (27.72 dB). } 
	\label{Fig:Deblurring results - barbara}
\end{figure*}

In our main evaluation, we examined the proposed Algorithms 2 and 3 for image deblurring in conjunction with the JPEG2000 and HEVC compression techniques (see the parameter settings in Table \ref{table:Experimental Results - Parameter Setting}). 
Table \ref{table:Deblurring - PSNR Comparison} shows a comparison between various deblurring methods tested in the above two settings for four grayscale images\footnote{The results in Table \ref{table:Deblurring - PSNR Comparison} for the methods from \cite{neelamani2004forward,guerrero2008image,dabov2008image,oliveira2009adaptive,chantas2010variational,danielyan2012bm3d} were taken as is from \cite{danielyan2012bm3d}.}. In 7 out of the 8 cases, the proposed Algorithms 2 or 3 utilizing the HEVC standard provided one of the best three results.
Visual results are presented in Figures \ref{Fig:Deblurring results - cameraman} and \ref{Fig:Deblurring results - barbara}.

\subsection{Image Inpainting}
\label{subsec:Inpainting}

We presented in \cite{dar2016image} experimental results for the inpainting problem, in its noisy and noiseless settings. 
Here we focus on the noiseless inpainting problem, where only pixel erasure occurs without an additive noise. 
The degradation is represented by a diagonal matrix $ \mtx{H} $ of $ N\times N $ size with main diagonal values of zeros and ones, indicating positions of missing and available pixels, respectively. Then, the product $ \mtx{H}\vec{x} $ equals to an $ N $-length vector where its $ k^{th} $ sample is determined by $ \mtx{H} $: if $ \mtx{H}[k,k] = 0$ then it is zero, and for $ \mtx{H}[k,k] = 1$ it equals to the corresponding sample of $ \vec{x} $.
The structure of the pixel erasure operator let us to simplify the optimization in step 6 of Algorithm \ref{Algorithm:Proposed Method Overlapping Blocks}. We note that $ \mtx{H} $ is a square diagonal matrix and, therefore, $ \mtx{H}^T = \mtx{H} $ and $ \mtx{H}^T \vec{y} $ is equivalent to a vector $ \vec{y} $ with zeroed components according to $ \mtx{H} $'s structure.
Additional useful relation is $ \mtx{H}^T \mtx{H} = \mtx{H} $.
Consequently, step 6 of Algorithm \ref{Algorithm:Proposed Method Overlapping Blocks} facilitates a componentwise computation that is interpreted to form the $ k^{th} $ sample of $  \hat{\vec{x}}^{(t)}  $ as 
\begin{IEEEeqnarray}{rCl}
	\label{eq:noiseless inpainting - k-th component of x}
	\hat{\vec{x}}^{(t)} [k] = \left\{
	\begin{array}{ll}
		\frac{ \vec{y}[k] + \frac{\beta}{2}  \sum\limits_{j=1}^{N_b}{{\tilde{\vec{z}}}^{j,(t)}[k]}  }{ 1 + \frac{\beta}{2} N_b }   ~~~~~~\text{for  }~\mtx{H}[k,k]=1\\
		\frac{1}{N_b} \sum\limits_{j=1}^{N_b} {\tilde{\vec{z}}}^{j,(t)}[k] ~~~~~~~~~ \text{for  }~\mtx{H}[k,k]=0 
	\end{array}~~
	\right.
\end{IEEEeqnarray}
We initialize the shifted images $ \left\lbrace {\hat{\vec{z}}}^{j,(0)} \right\rbrace _{j=1}^{N_b} $ as the given image with the missing pixels set as the corresponding local averages of the available pixels in the respective $ 7\times 7 $ neighborhoods.
When the iterative processing ends, we use the fact that the available pixels are noiseless and set them in the reconstructed image. The rest of the procedure remains as before.

\begin{figure*}[]
	\centering
	{\subfloat[{Original}]{\label{fig:inpainting - barbara - true image}\includegraphics[width=0.24\textwidth]{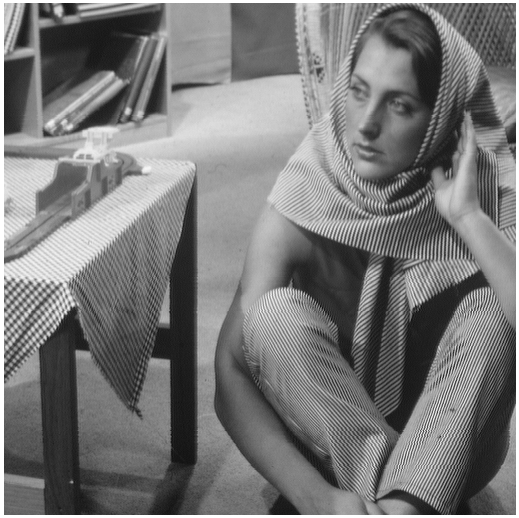}}}
	{\subfloat[{Deteriorated}]{\label{fig:inpainting - barbara - degraded image}\includegraphics[width=0.24\textwidth]{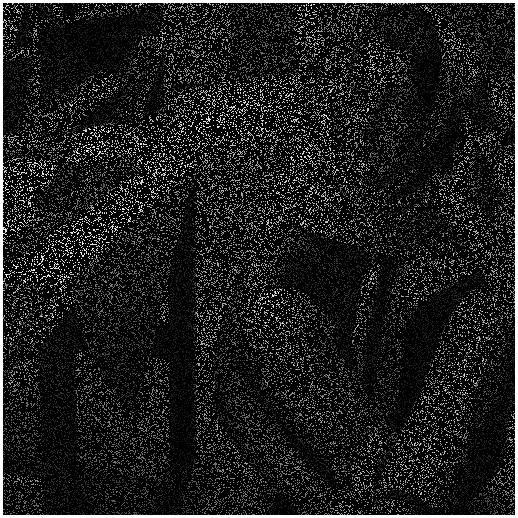}}}
	{\subfloat[{Proposed Restoration - JPEG2000}]{\label{fig:inpainting - barbara - proposed restoration - JPEG2000}\includegraphics[width=0.24\textwidth]{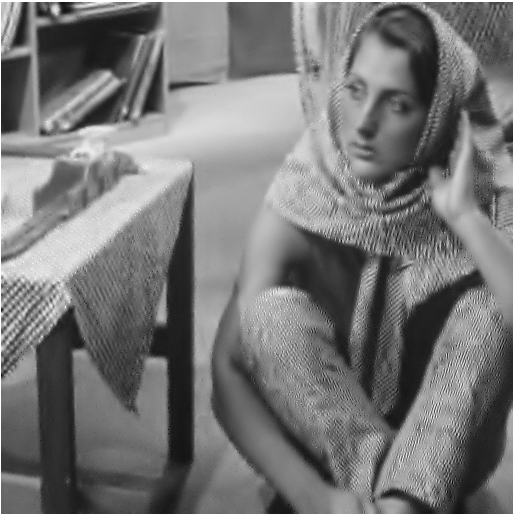}}}
	{\subfloat[{Proposed Restoration - HEVC}]{\label{fig:inpainting - barbara - proposed restoration}\includegraphics[width=0.24\textwidth]{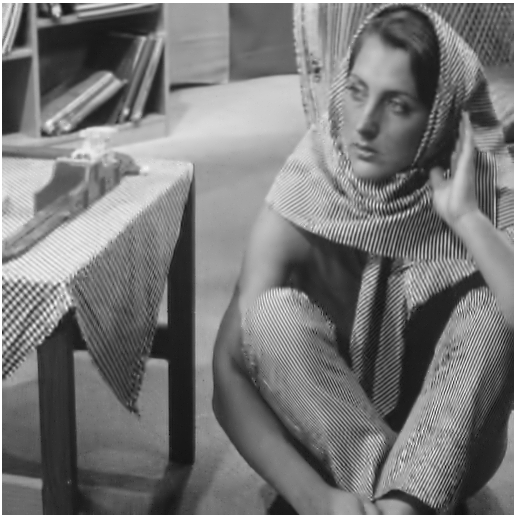}}}
	\caption{The inpainting experiment (80\% missing pixels) for the Barbara image ($ 512\times 512 $). (a) The original image. (b) Deteriorated image. (c) Restored image using Algorithm 3 with JPEG2000 compression (24.83 dB) (d) Restored image using Algorithm 3 with HEVC compression (28.80 dB). } 
	\label{Fig:Inpainting results - Barbara}
\end{figure*}

\begin{figure*}[]
	\centering
	{\subfloat[{Original}]{\label{fig:inpainting - house - true image}\includegraphics[width=0.24\textwidth]{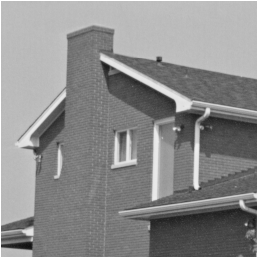}}}
	{\subfloat[{Deteriorated}]{\label{fig:inpainting - house - degraded image}\includegraphics[width=0.24\textwidth]{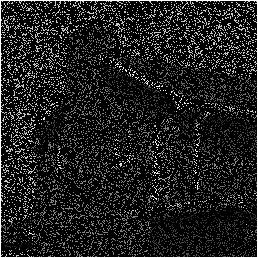}}}
	{\subfloat[{Proposed Restoration - JPEG2000}]{\label{fig:inpainting - house - proposed restoration - JPEG2000}\includegraphics[width=0.24\textwidth]{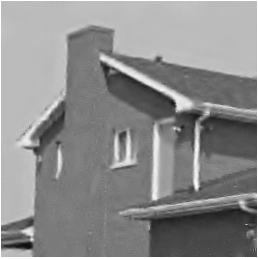}}}
	{\subfloat[{Proposed Restoration - HEVC}]{\label{fig:inpainting - house - proposed restoration}\includegraphics[width=0.24\textwidth]{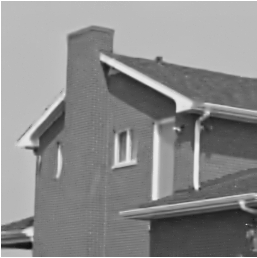}}}
	\caption{The inpainting experiment (80\% missing pixels) for the House image ($ 256\times 256 $). (a) The original image. (b) Deteriorated image. (c) Restored image using Algorithm 3 with JPEG2000 compression (30.50 dB)  (d) Restored image using Algorithm 3 with HEVC compression (33.10 dB). } 
	\label{Fig:Inpainting results - House}
\end{figure*}

We present here implementations of Algorithms \ref{Algorithm:Proposed Method Overlapping Blocks} and \ref{Algorithm:Proposed Method Overlapping Blocks with Robust Dual Variables} utilizing the JPEG2000 and HEVC image compression (the parameter settings are described in Table \ref{table:Experimental Results - Parameter Setting}).
We consider the experimental settings from \cite{ram2013image}, where 80\% of the pixels are missing (see Fig. \ref{fig:inpainting - barbara - degraded image} and \ref{fig:inpainting - house - degraded image}).
Five competing inpainting methods are considered: cubic interpolation of missing pixels via Delaunay triangulation (using Matlab's 'griddata' function); inpainting using sparse representations of patches of $ 16\times 16 $ pixels based on an overcomplete DCT (ODCT) dictionary (see method description in \cite[Ch. 15]{elad2010sparse}); using patch-group transformation \cite{li2008patch}; based on patch clustering \cite{yu2012solving}; and via patch reordering \cite{ram2013image}.
The PSNR values of images restored using the above methods (taken from \cite{ram2013image}) are provided in Table \ref{table:Experimental Results - Inpainting} together with our results. For two images our HEVC-based implementation of Algorithm 3 provides the highest PSNR values. Visually, Figures \ref{fig:inpainting - barbara - proposed restoration} and \ref{fig:inpainting - house - proposed restoration} exhibit the effectiveness of our method in repairing the vast amount of absent pixels.

\begin{table*} []
	\caption{Image Inpainting from 80\% Missing Pixels: PSNR Results}
	\renewcommand{\arraystretch}{1.1}
	\label{table:Experimental Results - Inpainting}
	\centering
		\begin{tabular}{|c||c|c|c|c|c|c|c|c|c|}
			\hline
			Image & \shortstack{Delaunay\\Triang.} &  ODCT & \shortstack{Li\\\cite{li2008patch}} & \shortstack{Yu et al.\\\cite{yu2012solving}} & \shortstack{Ram et al.\\\cite{ram2013image}} & \shortstack{\\Proposed\\Algorithm 2\\with JPEG2000} & \shortstack{\\Proposed\\Algorithm 3\\with JPEG2000} & \shortstack{\\Proposed\\Algorithm 2\\with HEVC} & \shortstack{\\Proposed\\Algorithm 3\\with HEVC} \\
			\hline\hline		                                       
			
			\shortstack{\\Lena\\512x512}  & 30.25 & 29.97 & 31.62 & 32.22 & 31.96 &  30.31 & 30.96 & 32.14 & \textbf{32.55} \\
			\hline		                                               		                       
			\shortstack{\\Barbara\\512x512}  & 22.88 & 27.15 & 25.40 & \textbf{30.94}  & 29.71 & 24.25 & 24.83 & 26.06 & 28.80 \\
			\hline		                                               					
			\shortstack{\\House\\256x256} & 29.21 & 29.69 & 32.87 & 33.05 & 32.71 &  29.49 & 30.50 & 32.42 & \textbf{33.10} \\
			\hline		                              
			                 		                       
		\end{tabular}
\end{table*}

\section{Conclusion}
\label{sec:Conclusion}
In this paper we explored the topic of complexity-regularized restoration, where the likelihood of candidate estimates are determined by their compression bit-costs. Using the alternating direction method of multipliers (ADMM) approach we developed three practical methods for restoration using standard compression techniques. Two of the proposed methods rely on a new shift-invariant complexity regularizer, evaluating the total bit-cost of the signal shifted versions. 
We explained few of the main ideas of our approach using an insightful theoretical-analysis of complexity-regularized restoration of a cyclo-stationary Gaussian signal from deterioration of a linear shift-invariant operator and additive white Gaussian noise. 
Experiments for deblurring and inpainting of images using the JPEG2000 and HEVC technique showed good results.


%

\appendix
\section{Proofs for the theory section}

\subsection{Equivalence of Problems \ref{problem: basic form} and \ref{problem: pseudoinverse filtered input}}
\label{appendix:Equivalence of Problems 1 and 2}

We start by showing the equality between the distortion constraints of Problems \ref{problem: basic form} and \ref{problem: pseudoinverse filtered input}. 
We develop the distortion of Problem \ref{problem: basic form} as follows: 
\begin{IEEEeqnarray}{rCl}
	\label{eq:appendix - proofs - equivalence of problem 1 and 2 - distortions}
	\left\| { \vec{y}  - \mtx{H} \hat{\vec{x}} } \right\|_2^2 & = & \left\| { \left( \mtx{I} - \mtx{H}\mtx{H}^{+} \right)\vec{y} + \mtx{H}\mtx{H}^{+} \vec{y} - \mtx{H} \hat{\vec{x}} } \right\|_2^2 
	\nonumber \\  
	 & = & \left\| { \left( \mtx{I} - \mtx{H}\mtx{H}^{+} \right)\vec{y} + \mtx{H} \left(  \mtx{H}^{+} \vec{y} - \hat{\vec{x}} \right) } \right\|_2^2 
	\nonumber \\  
	& = & \left\| {  \left( \mtx{I} - \mtx{H}\mtx{H}^{+} \right) \vec{y} } \right\|_2^2  + \left\| {  \mtx{H} \left( \mtx{H}^{+} \vec{y} -  \hat{\vec{x}} \right) } \right\|_2^2 \nonumber\\ 
	&& + \left( \mtx{H}^{+} \vec{y} -  \hat{\vec{x}} \right)^{*} \mtx{H}^{*} \left( \mtx{I} - \mtx{H}\mtx{H}^{+} \right) \vec{y} \nonumber\\ 
	&& + \vec{y}^{*} \left( \mtx{I} - \mtx{H}\mtx{H}^{+} \right)^{*} \mtx{H} \left(  \mtx{H}^{+} \vec{y} - \hat{\vec{x}} \right)
	\nonumber \\  
	& = & \left\| {  \left( \mtx{I} - \mtx{H}\mtx{H}^{+} \right) \vec{y} } \right\|_2^2  + \left\| {  \mtx{H} \left( \mtx{H}^{+} \vec{y} -  \hat{\vec{x}} \right) } \right\|_2^2  ~~~~
\end{IEEEeqnarray}
where the last equality follows from 
\begin{IEEEeqnarray}{rCl}
	\label{eq:appendix - proofs - equivalence of problem 1 and 2 - distortions - auxiliary result}
	\mtx{H}^{*}  \left( \mtx{I} - \mtx{H}\mtx{H}^{+} \right)  = \mtx{0} 
\end{IEEEeqnarray}
that can be easily proved, e.g., by using the DFT-based diagonalization of $ \mtx{H} $ and $ \mtx{H}^{+} $.

The first term in (\ref{eq:appendix - proofs - equivalence of problem 1 and 2 - distortions}) can be further developed: 
\begin{IEEEeqnarray}{rCl}
	\label{eq:appendix - proofs - equivalence of problem 1 and 2 - distortions - first part - deterministic}
	\left\| {  \left( \mtx{I} - \mtx{H}\mtx{H}^{+} \right) \vec{y} } \right\|_2^2  & = & \left\| {  \left( \mtx{I} - \mtx{F}^{*}\mtx{\Lambda}_H \mtx{F} \mtx{F}^{*}\mtx{\Lambda}_H^{+} \mtx{F} \right) \vec{y} } \right\|_2^2 \nonumber \\
	& = & \left\| { \mtx{F}^{*} \left( \mtx{I} - \mtx{\Lambda}_H \mtx{\Lambda}_H^{+}  \right)\mtx{F} \vec{y} } \right\|_2^2 	\nonumber \\
	& = & \left\| { \left( \mtx{I} - \mtx{\Lambda}_H \mtx{\Lambda}_H^{+}  \right) \vec{y}^{F}  } \right\|_2^2
	\nonumber \\
	& = & \sum\limits_{k: h_k^F = 0} \left\lvert y_k^F  \right\rvert ^2 = \nonumber \\
	& = & \sum\limits_{k: h_k^F = 0} \left\lvert n_k^F  \right\rvert ^2
\end{IEEEeqnarray}
where $ \vec{y}^{F} \triangleq \mtx{F} \vec{y} $ is the DFT-domain representation of $ \vec{y} $ (correspondingly, we use these notations to any vector), and the last equality is implied from the DFT-component relation $ y_k^F = h_k^F x_k^F + n_k^F $ that reduces to $ y_k^F = n_k^F $ for components with $ h_k^F = 0 $. 
Consequently, 
\begin{IEEEeqnarray}{rCl}
	\label{eq:appendix - proofs - equivalence of problem 1 and 2 - distortions - first part}
	E\left\lbrace \left\| {  \left( \mtx{I} - \mtx{H}\mtx{H}^{+} \right) \vec{y} } \right\|_2^2  \right\rbrace & = & E\left\lbrace  \sum\limits_{k: h_k^F = 0} \left\lvert n_k^F  \right\rvert ^2 \right\rbrace 	\nonumber \\
	& = &   (N - N_{H}) \sigma_n^2 
\end{IEEEeqnarray}
where $ N_{H} $ was defined in Section \ref{subsec:Reformulations of the Problem} as the rank of $ \mtx{H} $. 
Accordingly, and also using (\ref{eq:appendix - proofs - equivalence of problem 1 and 2 - distortions}), the distortion constraint of Problem \ref{problem: basic form}, i.e., 
\begin{IEEEeqnarray}{rCl}
	\label{eq:appendix - proofs - equivalence of problem 1 and 2 - distortion of problem 1}
	E\left\lbrace \left\| { \vec{y}  - \mtx{H} \hat{\vec{x}} } \right\|_2^2  \right\rbrace = N \sigma_n^2
\end{IEEEeqnarray} 
equals to (recall that $ \tilde{\vec{y}} = \mtx{H}^{+} \vec{y} $)
\begin{IEEEeqnarray}{rCl}
	\label{eq:appendix - proofs - equivalence of problem 1 and 2 - distortion of problem 2}
	E\left\lbrace \left\| \mtx{H} \left(  { {\tilde{\vec{y}}}  -  \hat{\vec{x}} } \right) \right\|_2^2  \right\rbrace = N_{H} \sigma_n^2 , 
\end{IEEEeqnarray} 
that is, the distortion constraint of Problem \ref{problem: pseudoinverse filtered input}.

We now turn to prove the equivalence of Problems \ref{problem: basic form} and \ref{problem: pseudoinverse filtered input}. Our proof sketch conforms with common arguments in rate-distortion function proofs (see \cite{cover2012elements}): first, we lower bound the mutual information $ I\left( \vec{y}; \hat{\vec{x}} \right) $, which is the cost function of Problem \ref{problem: basic form}; then, we provide a statistical construction achieving the lower bound while obeying the distortion constraint.

The proposed lower bound for $ I\left( \vec{y}; \hat{\vec{x}} \right) $ is established by noting that $ \tilde{\vec{y}} = \mtx{H}^{+} {\vec{y}} $ and, therefore, the data processing inequality \cite{cover2012elements} implies here that 
\begin{IEEEeqnarray}{rCl}
	\label{eq:appendix - proofs - equivalence of problem 1 and 2 - data processing inequality}
	I\left( \vec{y}; \hat{\vec{x}} \right) \ge I\left( \tilde{\vec{y}} ; \hat{\vec{x}} \right) , 
\end{IEEEeqnarray} 
where $ I\left( \tilde{\vec{y}} ; \hat{\vec{x}} \right) $ is the cost function of Problem \ref{problem: pseudoinverse filtered input}.
The relation in (\ref{eq:appendix - proofs - equivalence of problem 1 and 2 - data processing inequality}) is known to be attained with equality when  $ \vec{y} $ and $\hat{\vec{x}} $ are independent given $ \tilde{\vec{y}} $. The next construction shows that this is indeed the case. 

We will now show the achievability of the lower bound in (\ref{eq:appendix - proofs - equivalence of problem 1 and 2 - data processing inequality}) by describing a two-stage setting that statistically represents $ \tilde{\vec{y}} $ as an outcome of $ \hat{\vec{x}} $, and $ {\vec{y}} $ as a consequence of $ \tilde{\vec{y}} $. This layout is an instance of the construction concept known as the (backward) test channel \cite{cover2012elements}.
The first stage of our construction is based on 
\begin{IEEEeqnarray}{rCl}
	\label{eq:appendix - proofs - equivalence of problem 1 and 2 - backward channel construction - x }
	\hat{\vec{x}} & \sim & \mathcal{N} \left( \vec{0}, \mtx{H}^{+} \mtx{R}_{\vec{y}} \mtx{H}^{+*} - \sigma_n^2 \mtx{H}^{+}\mtx{H}^{+*} \right) 
	\\ 
	\label{eq:appendix - proofs - equivalence of problem 1 and 2 - backward channel construction - z }
	\vec{z} & \sim & \mathcal{N} \left( \vec{0}, \sigma_n^2 \mtx{H}^{+}\mtx{H}^{+*} \right) 
\end{IEEEeqnarray} 
where $ \hat{\vec{x}} $ and $ \vec{z} $ are independent. Consequently, we define 
\begin{IEEEeqnarray}{rCl}
	\label{eq:appendix - proofs - equivalence of problem 1 and 2 - backward channel construction - y_tilde as a sum }
	\tilde{\vec{y}} = \hat{\vec{x}} + \vec{z} , 
\end{IEEEeqnarray} 
implying $ \tilde{\vec{y}}  \sim \mathcal{N} \left( \vec{0}, \mtx{H}^{+} \mtx{R}_{\vec{y}} \mtx{H}^{+*} \right) $ that, indeed, conforms with $ \tilde{\vec{y}} = \mtx{H}^{+} \vec{y} $ where $\vec{y} \sim \mathcal{N} \left( \vec{0}, \mtx{R}_{\vec{y}} \right)$. 
Moreover, the construction (\ref{eq:appendix - proofs - equivalence of problem 1 and 2 - backward channel construction - x })-(\ref{eq:appendix - proofs - equivalence of problem 1 and 2 - backward channel construction - y_tilde as a sum }) yields 

\begin{IEEEeqnarray}{rCl}
	\label{eq:appendix - proofs - equivalence of problem 1 and 2 - backward channel construction - distortion constraint satisfies}
	E\left\lbrace \left\| \mtx{H} \left(  { {\tilde{\vec{y}}}  -  \hat{\vec{x}} } \right) \right\|_2^2  \right\rbrace & = & E\left\lbrace \left\| \mtx{H} \vec{z} \right\|_2^2  \right\rbrace \nonumber
	\\
	& = & E\left\lbrace \vec{z}^{*}  \mtx{H}^{*}  \mtx{H} \vec{z} \right\rbrace \nonumber\\
	& = & E\left\lbrace Trace\left\lbrace \vec{z}^{*}  \mtx{H}^{*}  \mtx{H} \vec{z} \right\rbrace \right\rbrace \nonumber\\
	& = & E\left\lbrace Trace\left\lbrace \mtx{H} \vec{z} \vec{z}^{*}  \mtx{H}^{*}   \right\rbrace \right\rbrace \nonumber\\
	& = & Trace\left\lbrace \mtx{H} \mtx{R}_{\vec{z}} \mtx{H}^{*}   \right\rbrace 
	\nonumber\\
	& = & \sigma_n^2 \cdot Trace\left\lbrace \mtx{H}  \mtx{H}^{+}\mtx{H}^{+*} \mtx{H}^{*}   \right\rbrace 
	\nonumber\\
	& = & \sigma_n^2 \cdot Trace\left\lbrace \mtx{P}_{{H}}  \mtx{P}_{{H}}^{*}   \right\rbrace 
	\nonumber\\
	& = & \sigma_n^2 N_H
\end{IEEEeqnarray} 
where $ \mtx{P}_{{H}} \triangleq \mtx{H}  \mtx{H}^{+}  $ is the matrix projecting onto the range of $ \mtx{H} $, note it is also a circulant matrix diagonalized by the DFT matrix to the diagonal matrix $ \mtx{\Lambda}_{\mtx{P}_{{H}}} \triangleq \mtx{\Lambda}_{H} \mtx{\Lambda}_{H}^{+}  $. The last computation of the trace is due to the structure of $\mtx{\Lambda}_{\mtx{P}_{H}}$, having ones in main-diagonal entries corresponding to the DFT-domain indices of the range of $ \mtx{H} $, and zeros elsewhere. 
The result in (\ref{eq:appendix - proofs - equivalence of problem 1 and 2 - backward channel construction - distortion constraint satisfies}) shows that the distortion constraint (\ref{eq:appendix - proofs - equivalence of problem 1 and 2 - distortion of problem 2}) is satisfied.

Let us consider the second stage of the construction, awaiting to prove that $ \vec{y} $ and $\hat{\vec{x}} $ are independent given $ \tilde{\vec{y}} $. 
We precede the construction with examining the following decomposition of $ \vec{y} $ 
\begin{IEEEeqnarray}{rCl}
	\label{eq:appendix - proofs - equivalence of problem 1 and 2 - backward channel construction - decomposition of y}
	\vec{y} & = & \mtx{P}_{H}  \vec{y} + \left( \mtx{I} - \mtx{P}_{H} \right) \vec{y}
	\nonumber \\ \nonumber
	& = & \mtx{H} \mtx{H}^{+}  \vec{y} + \left( \mtx{I} - \mtx{P}_{H} \right) \left( \mtx{H} \vec{x} + \vec{n}\right) 
	\\ 
	& = & \mtx{H} \tilde{\vec{y}} + \left( \mtx{I} - \mtx{P}_{H} \right) \vec{n}
	\end{IEEEeqnarray} 
where the second equality uses the degradation model, and the third equality is due to $ \left( \mtx{I} - \mtx{P}_{H} \right) \mtx{H} = \mtx{0} $. 
Importantly, Eq. (\ref{eq:appendix - proofs - equivalence of problem 1 and 2 - backward channel construction - decomposition of y}) describes $ \vec{y} $ as a linear combination of two independent random vectors: $ \tilde{\vec{y}} $ and $ \left( \mtx{I} - \mtx{P}_{H} \right) \vec{n} $. Since  $ \tilde{\vec{y}} $ and $ \left( \mtx{I} - \mtx{P}_{H} \right) \vec{n} $ are Gaussian random vectors, their independence is proved by showing they are uncorrelated via  
\begin{IEEEeqnarray}{rCl}
	\label{eq:appendix - proofs - equivalence of problem 1 and 2 - backward channel construction - uncorrelated backward variables - new }
	&& E\left\lbrace  \left( \mtx{I} - \mtx{P}_{H} \right) \vec{n} \tilde{\vec{y}}^{*} \right\rbrace 
	\nonumber \\ \nonumber
	&& = E\left\lbrace  \left( \mtx{I} - \mtx{P}_{H} \right) \vec{n} \vec{y}^{*} \mtx{H}^{+*} \right\rbrace 
	\\ \nonumber
	&& = E\left\lbrace  \left( \mtx{I} - \mtx{F}^{*} \mtx{\Lambda}_{\mtx{P}_{H}} \mtx{F} \right) \vec{n} \vec{y}^{*} \mtx{F}^{*} \mtx{\Lambda}_{H}^{+*} \mtx{F} \right\rbrace 
	\\ \nonumber
	&& = E\left\lbrace  \mtx{F}^{*} \left( \mtx{I} -  \mtx{\Lambda}_{\mtx{P}_{H}}  \right) \mtx{F} \vec{n} \left( \mtx{F} \vec{y} \right)^{*}  \mtx{\Lambda}_{H}^{+*} \mtx{F} \right\rbrace
	\\ \label{eq:appendix - proofs - equivalence of problem 1 and 2 - backward channel construction - uncorrelated backward variables - new - explain}
	&& = \mtx{F}^{*} E\left\lbrace   \left( \mtx{I} -  \mtx{\Lambda}_{\mtx{P}_{H}}  \right) \vec{n}^{F} \left( \mtx{\Lambda}_{H}^{+} \vec{y}^{F} \right)^{*}   \right\rbrace  \mtx{F}
	\\ \nonumber
	&& = \mtx{F}^{*} \mtx{0}  \mtx{F}
	\\ \nonumber
	&& = \mtx{0}
\end{IEEEeqnarray} 
where in (\ref{eq:appendix - proofs - equivalence of problem 1 and 2 - backward channel construction - uncorrelated backward variables - new - explain}) we used the facts that $\left( \mtx{I} -  \mtx{\Lambda}_{\mtx{P}_{H}}  \right) \vec{n}^{F}$ is a DFT-domain vector with zeros in components corresponding to the range of $ \mtx{H} $, and $ \mtx{\Lambda}_{H}^{+} \vec{y}^{F} $ is a DFT-domain vector with zeros in entries corresponding to the nullspace of $ \mtx{H} $, hence, these zero patterns yield the outer-product matrix which is all zeros. 

The decomposition in (\ref{eq:appendix - proofs - equivalence of problem 1 and 2 - backward channel construction - decomposition of y}) motivates us to consider $ \vec{y} $ to emerge from $ \tilde{\vec{y}} $ via the statistical relation 
\begin{IEEEeqnarray}{rCl}
	\label{eq:appendix - proofs - equivalence of problem 1 and 2 - backward channel construction - y from y_tilde }
	\vec{y} = \mtx{H} \tilde{\vec{y}} + \vec{w}
\end{IEEEeqnarray} 
where $ \vec{w} \sim \mathcal{N} \left( \vec{0}, \sigma_n^2  \left(\mtx{I} - \mtx{P}_{H}  \right) \left(\mtx{I} - \mtx{P}_{H}  \right)^{*} \right) $ and is independent of $ \tilde{\vec{y}} $ (and also of $ \vec{z} $). Note that $ \vec{w} $ takes the role of $ \left( \mtx{I} - \mtx{P}_{H} \right) \vec{n} $ appearing in (\ref{eq:appendix - proofs - equivalence of problem 1 and 2 - backward channel construction - decomposition of y}), e.g., they have the same distribution.
One can further examine the suitability of the construction (\ref{eq:appendix - proofs - equivalence of problem 1 and 2 - backward channel construction - y from y_tilde }) to the considered problem by noting it satisfies the distortion constraint of Problem \ref{problem: basic form}, namely, 
\begin{IEEEeqnarray}{rCl}
	\label{eq:appendix - proofs - equivalence of problem 1 and 2 - backward channel construction - distortion constraint satisfies - Problem 2}
	E\left\lbrace \left\| { \vec{y}  - \mtx{H} \hat{\vec{x}} } \right\|_2^2 \right\rbrace & = &  E\left\lbrace \left\| { \mtx{H} \tilde{\vec{y}} + \vec{w}  - \mtx{H} \hat{\vec{x}} } \right\|_2^2 \right\rbrace
	\\ \nonumber
	& = &  E\left\lbrace \left\| { \mtx{H} \left( \hat{\vec{x}} + \vec{z} \right) + \vec{w}  - \mtx{H} \hat{\vec{x}} } \right\|_2^2 \right\rbrace
	\\ \nonumber
	& = &  E\left\lbrace \left\| { \mtx{H} \vec{z} + \vec{w} } \right\|_2^2 \right\rbrace
	\\ \nonumber
	& = &  E\left\lbrace \left\| { \mtx{H} \vec{z} } \right\|_2^2 \right\rbrace + E\left\lbrace \left\| {  \vec{w} } \right\|_2^2 \right\rbrace
	\\ \nonumber
	& = &  \sigma_n^2 Trace\left\lbrace \mtx{H}  \mtx{H}^{+}\mtx{H}^{+*} \mtx{H}^{*}  \right\rbrace  \\ \nonumber 
	&& +  \sigma_n^2 Trace\left\lbrace \left(\mtx{I} - \mtx{P}_{H}  \right) \left(\mtx{I} - \mtx{P}_{H}  \right)^{*} \right\rbrace
	\\ \nonumber
	& = &  \sigma_n^2 Trace\left\lbrace \mtx{P}_{H} \mtx{P}_{H}^{*}  \right\rbrace  \\ \nonumber 
	&& +  \sigma_n^2 Trace\left\lbrace \left(\mtx{I} - \mtx{P}_{H}  \right) \left(\mtx{I} - \mtx{P}_{H}  \right)^{*} \right\rbrace
	\\ \nonumber
	& = &  \sigma_n^2 \cdot N_H + \sigma_n^2 \cdot \left( N - N_H \right)
	\\ \nonumber
	& = &  N \sigma_n^2 
\end{IEEEeqnarray} 
as required. 
Furthermore, the constructed $ \vec{y} $ indeed obeys  $\vec{y} \nolinebreak \sim \nolinebreak \mathcal{N} \left( \vec{0}, \mtx{R}_{\vec{y}} \right)$. Specifically, its autocorrelation matrix stems from the calculation
\begin{IEEEeqnarray}{rCl}
	\label{eq:appendix - proofs - equivalence of problem 1 and 2 - backward channel construction - Problem 2 - y autocorrelation}
	E\left\lbrace \vec{y} \vec{y}^{*} \right\rbrace & = & E\left\lbrace  \left( { \mtx{H} \tilde{\vec{y}} + \vec{w} }  \right) \left( { \mtx{H} \tilde{\vec{y}} + \vec{w} }  \right)^{*} \right\rbrace
	\\ \nonumber
	 & = & E\left\lbrace  \left( { \mtx{H} \tilde{\vec{y}} }  \right) \left( { \mtx{H} \tilde{\vec{y}}  }  \right)^{*} \right\rbrace +  E\left\lbrace   { \vec{w} }  { \vec{w} }^{*} \right\rbrace
	 \\ \nonumber
	 & = & \mtx{H} \mtx{H}^{+} \mtx{R}_{\vec{y}} \mtx{H}^{+*} \mtx{H}^{*} +  \mtx{R}_{\vec{w}}
	 \\ \nonumber
	 & = & \mtx{P}_{H} \left( \mtx{H} \mtx{R}_{\vec{x}} \mtx{H}^{*} + \sigma_n^2 \mtx{I}  \right) \mtx{P}_{H}^{*} +  \mtx{R}_{\vec{w}}
	 \\ \nonumber
	 & = & \mtx{P}_{H} \mtx{H} \mtx{R}_{\vec{x}} \mtx{H}^{*} \mtx{P}_{H}^{*} + \sigma_n^2 \mtx{P}_{H} \mtx{P}_{H}^{*} +  \mtx{R}_{\vec{w}}
	 \\ \nonumber
	 & = & \mtx{H} \mtx{R}_{\vec{x}} \mtx{H}^{*} + \sigma_n^2 \mtx{P}_{H} \mtx{P}_{H}^{*} +  \sigma_n^2  \left(\mtx{I} - \mtx{P}_{H}  \right) \left(\mtx{I} - \mtx{P}_{H}  \right)^{*}
	 \\ \nonumber
	 & = & \mtx{H} \mtx{R}_{\vec{x}} \mtx{H}^{*} + \sigma_n^2 \mtx{I} 
	 \\ \nonumber
	 & = & \mtx{R}_{\vec{y}}  
\end{IEEEeqnarray}
where we used the auxiliary result 
\begin{IEEEeqnarray}{rCl}
	\label{eq:appendix - proofs - equivalence of problem 1 and 2 - backward channel construction - Problem 2 - y autocorrelation - auxiliary result}
	 && \mtx{P}_{H} \mtx{P}_{H}^{*} + \left(\mtx{I} - \mtx{P}_{H}  \right) \left(\mtx{I} - \mtx{P}_{H}  \right)^{*}
	 \\ \nonumber
	&& = \mtx{F}^{*} \mtx{\Lambda}_{\mtx{P}_{H}} \mtx{\Lambda}_{\mtx{P}_{H}}^{*} \mtx{F} + \mtx{F}^{*} \left(\mtx{I} - \mtx{\Lambda}_{\mtx{P}_{H}}   \right) \left(\mtx{I} - \mtx{\Lambda}_{\mtx{P}_{H}}  \right)^{*} \mtx{F} 
	\\ \nonumber
	&& = \mtx{F}^{*} \mtx{\Lambda}_{\mtx{P}_{H}} \mtx{F} + \mtx{F}^{*} \left(\mtx{I} - \mtx{\Lambda}_{\mtx{P}_{H}}   \right)  \mtx{F} 
	\\ \nonumber
	&& = \mtx{I} .
\end{IEEEeqnarray}

Joining the two stages of the construction, presented in (\ref{eq:appendix - proofs - equivalence of problem 1 and 2 - backward channel construction - y_tilde as a sum }) and (\ref{eq:appendix - proofs - equivalence of problem 1 and 2 - backward channel construction - y from y_tilde }), exhibits $ \hat{\vec{x}} \rightarrow \tilde{\vec{y}} \rightarrow \vec{y} $ as a Markov chain and, therefore, $ \vec{y} $ and $\hat{\vec{x}} $ are independent given $ \tilde{\vec{y}} $. 
This evident construction turns (\ref{eq:appendix - proofs - equivalence of problem 1 and 2 - data processing inequality}) into 
\begin{IEEEeqnarray}{rCl}
	\label{eq:appendix - proofs - equivalence of problem 1 and 2 - mutual information equality}
	I\left( \vec{y}; \hat{\vec{x}} \right) = I\left( \tilde{\vec{y}} ; \hat{\vec{x}} \right) 
\end{IEEEeqnarray} 
that completes proving the equivalence of Problems \ref{problem: basic form} and \ref{problem: pseudoinverse filtered input}.

\subsection{Equivalence of Problems \ref{problem: pseudoinverse filtered input} and \ref{problem: Separable Form in DFT domain}}
\label{appendix:Equivalence of Problems 2 and 3}

The rate-distortion function for a Gaussian source with memory (i.e., correlated components) is usually derived in the Principle Component Analysis (PCA) domain where the components are independent Gaussian variables (see, e.g., \cite{berger1971rate}). In our case, where the signal is cyclo-stationary, the PCA is obtained using the DFT matrix. As in the usual case, 
\begin{IEEEeqnarray}{rCl}
	\label{eq:appendix - proofs - equivalence of problem 2 and 3 - mutual information equality under DFT}
	I\left( \tilde{\vec{y}} ; \hat{\vec{x}} \right) & = & I\left( \tilde{\vec{y}}^{F} ; \hat{\vec{x}}^{F} \right) 
	\\ 
	\label{eq:appendix - proofs - equivalence of problem 2 and 3 - mutual information reduction to sum}
	& = &  \sum\limits_{k=0 }^{N-1} I\left( {\tilde{y}^F_k} ; \hat{x}^F_k \right)
\end{IEEEeqnarray}
where (\ref{eq:appendix - proofs - equivalence of problem 2 and 3 - mutual information equality under DFT}) emerges from the reversibility of the transformation, and (\ref{eq:appendix - proofs - equivalence of problem 2 and 3 - mutual information reduction to sum}) is due to the independence of the $ \left\lbrace {\tilde{y}^F_k} \right\rbrace_{k=0}^{N-1} $ components \cite{cover2012elements} and the backward-channel construction (see Appendix \ref{appendix:Equivalence of Problems 1 and 2}) that can be translated into a form of independent DFT-domain component-level channels.

The main difference from the well-known rate-distortion analysis is that here, in Problem \ref{problem: pseudoinverse filtered input}, the distortion constraint is not a regular squared error -- but a weighted one, that will be developed next. 
Since DFT is a unitary transformation, its energy preservation property yields 
\begin{IEEEeqnarray}{rCl}
	\label{eq:appendix - proofs - equivalence of problem 2 and 3 - distortion energy preservation}
	E\left\lbrace \left\| \mtx{H} \left(  { {\tilde{\vec{y}}}  -  \hat{\vec{x}} } \right) \right\|_2^2  \right\rbrace & = & E\left\lbrace \left\| \mtx{\Lambda}_{H} \left(  { {\tilde{\vec{y}}}^F  -  \hat{\vec{x}}^F } \right) \right\|_2^2 \right\rbrace
	\\ \nonumber 
	& = & \sum\limits_{k=0}^{N-1} { \left| h^F_k \right|^2 E\left\lbrace \left| {  \tilde{y}^F_k - \hat{x}^F_k } \right|^2 \right\rbrace }
\end{IEEEeqnarray} 
where the last equality is due to the diagonal structure of $ \mtx{\Lambda}_{H} $. Hence, we got that the two expected-distortion expressions in Problems \ref{problem: pseudoinverse filtered input} and \ref{problem: Separable Form in DFT domain} are equal.

\subsection{Solution of Problem \ref{problem: Gaussian distortion allocation}}
\label{appendix:Solution of Problem 4}

For a start, the transition between Problem \ref{problem: Separable Form in DFT domain} and Problem \ref{problem: Gaussian distortion allocation} is analogous to the familiar case of jointly coding a set of independent Gaussian variables \cite{cover2012elements}. Accordingly, and also due to lack of space, we do not elaborate here on this problem-equivalence proof. 

Problem \ref{problem: Gaussian distortion allocation} is compelling as it is a distortion-allocation optimization, where the distortion levels $ \left\lbrace D_k \right\rbrace_{k=0}^{N-1} $  are allocated under the joint distortion constraint. 
We address Problem \ref{problem: Gaussian distortion allocation} via its Lagrangian form (temporarily ignoring the constraints of non-negative distortions)
\begin{IEEEeqnarray}{rCl}
	\label{eq:appendix - proofs - solution of problem 4 - Lagrangian form}
	\underset{D_0,...,D_{N-1}}{\text{min}}
	\sum\limits_{k=0}^{N-1} { \left[ \frac{1}{2} \log\left( \frac{\lambda^{\left(\tilde{\vec{y}}_k \right) }}{D_k} \right) \right]_{+} }  + \mu \sum\limits_{k=0}^{N-1} { \left| h^F_k \right|^2 D_k  } \nonumber\\
\end{IEEEeqnarray}
where $ \mu \ge 0 $ is the Lagrange multiplier. 
Recalling that some components may correspond to $ h^F_k = 0 $ and, by (\ref{eq:theoretic Gaussian analysis - pseudoinverse filtered input - autocorrelation eigenvalues}), also $\lambda^{\left(\tilde{\vec{y}}\right)}_k = 0$ -- meaning they are deterministic variables. 
These deterministic components do not need to be coded (i.e., $ R_k \nolinebreak = \nolinebreak 0 $) while still attaining $ D_k = 0 $.
Accordingly, the Lagrangian optimization (\ref{eq:appendix - proofs - solution of problem 4 - Lagrangian form}) is updated into 
\begin{IEEEeqnarray}{rCl}
	\label{eq:appendix - proofs - solution of problem 4 - Lagrangian form - only components with nonzero h_k^F}
	\underset{ \left\lbrace D_k \right\rbrace_{k: h^F_k \ne 0} }{\text{min}}
	\sum\limits_{k: h^F_k \ne 0} { \frac{1}{2} \log\left( \frac{\lambda^{(\vec{x})}_k + \frac{\sigma_n^2}{\left| h^F_k \right|^2}}{D_k} \right) }  + \mu \sum\limits_{k: h^F_k \ne 0} { \left| h^F_k \right|^2 D_k  } \nonumber\\
\end{IEEEeqnarray}
where we used the expression from (\ref{eq:theoretic Gaussian analysis - pseudoinverse filtered input - autocorrelation eigenvalues}), and assumed that the distortions are small enough such that the operator $ \left[\cdot\right]_+ $ can be omitted (a correct assumption as will be later shown).
Now, the optimal $ D_k $ value can be determined by equating the respective derivative of the Lagrangian cost to zero, leading to allocated distortion (still as a function of $ \mu $) 
\begin{IEEEeqnarray}{rCl}
	\label{eq:appendix - proofs - solution of problem 4 - optimal distortion allocation - function of mu}
	D_k^{opt} = \frac{1}{2\ln{(2)}\mu \left| h^F_k \right|^2 }   \text{~~~for~} k: h^F_k \ne 0
\end{IEEEeqnarray}
and by setting the $ \mu $ satisfying the total distortion constraint from Problem \ref{problem: Gaussian distortion allocation} we get 
\begin{IEEEeqnarray}{rCl}
	\label{eq:appendix - proofs - solution of problem 4 - optimal distortion allocation}
	D_k^{opt} = \frac{\sigma_n^2}{\left| h^F_k \right|^2 }   \text{~~~for~} k: h^F_k \ne 0.
\end{IEEEeqnarray}
Expressing a nonuniform distortion-allocation (for components with nonzero $ h^F_k $), being inversely proportional to the weights $ \{ \left| h^F_k \right|^2 \} _{k=0}^{N-1} $.  One should note that the assumption on small-enough distortions is satisfied as $ D_k^{opt} \le \lambda^{\left(\tilde{\vec{y}}\right) }_k $ for any $ k $ obeying $ h^F_k \ne 0 $, and that all the distortions are non-negative as required. 
The optimal distortions established here (for $ k $ where $ h^F_k \ne 0 $) are set in the rate formula (\ref{eq:theoretic Gaussian analysis - rate-distortion function of a Gaussian variable}), providing the optimal rate allocation 
\begin{IEEEeqnarray}{rCl}
	\label{eq:appendix - proofs - solution of problem 4 - optimal rate allocation}
	R_k^{opt} = \left\{ {\begin{array}{*{20}{c}}
			{\frac{1}{2} \log\left( \left| h^F_k \right|^2 \frac{\lambda^{(\vec{x})}_k  }{\sigma_n^2} + 1 \right)}&{,\text{for~~} h^F_k\ne 0 } \\ 
			0&{,\text{for~~} h^F_k = 0 . } 
		\end{array}} \right.  	
\end{IEEEeqnarray}

\subsection{Equivalent Form of Stage 6 of Algorithm \ref{Algorithm:Proposed Method Non-overlapping}}
\label{appendix:Equivalent Forms of Stage 4 of Algorithm 1}

The analytic solution of Stage 6 of Algorithm \ref{Algorithm:Proposed Method Non-overlapping} is considered here with the conjugate-transpose operator, $ ^* $, extending the regular transpose:
\begin{IEEEeqnarray}{rCl}
	\label{eq:appendix - proofs - Equivalent Forms of Stage 4 of Algorithm 1 - signal domain}
	\hat{ \vec{x}}^{(t)} & = & \left(  \mtx{H}^{*}\mtx{H} + \frac{\beta}{2} \mtx{I}  \right)^{-1} \left( \mtx{H}^* \vec{y} + \frac{\beta}{2} \tilde{\vec{z}}^{(t)}   \right) .
\end{IEEEeqnarray}
We note that 
\begin{IEEEeqnarray}{rCl}
	\label{eq:appendix - proofs - Equivalent Forms of Stage 4 of Algorithm 1 - H^*y}
	\mtx{H}^* \vec{y} & = & \mtx{H}^* \left( \mtx{H} \mtx{H}^{+} \vec{y} + \left(\mtx{I} - \mtx{H} \mtx{H}^{+} \right) \vec{y} \right) \nonumber 
	\\
	& = & \mtx{H}^{*} \mtx{H} \tilde{\vec{y}} + \mtx{H}^{*} \left(\mtx{I} - \mtx{H} \mtx{H}^{+} \right) \vec{y}  \nonumber 
	\\
	& = & \mtx{H}^{*} \mtx{H} \tilde{\vec{y}}
\end{IEEEeqnarray}
where the last equality results from the relation $ \mtx{H}^{*} \left(\mtx{I} - \mtx{H} \mtx{H}^{+} \right) = \mtx{0} $. Consequently, (\ref{eq:appendix - proofs - Equivalent Forms of Stage 4 of Algorithm 1 - signal domain}) becomes 
\begin{IEEEeqnarray}{rCl}
	\label{eq:appendix - proofs - Equivalent Forms of Stage 4 of Algorithm 1 - signal domain - 2}
	\hat{ \vec{x}}^{(t)} & = & \left(  \mtx{H}^{*}\mtx{H} + \frac{\beta}{2} \mtx{I}  \right)^{-1} \left(  \mtx{H}^{*} \mtx{H} \tilde{\vec{y}} + \frac{\beta}{2} \tilde{\vec{z}}^{(t)}   \right) , 
\end{IEEEeqnarray}
which is the form presented in (\ref{eq:theory-practice relation - interpreting first stage}).

\ifCLASSOPTIONcaptionsoff
  \newpage
\fi



\bibliographystyle{IEEEtran}
\bibliography{IEEEabrv,complexity_regularized_inverse_problems__refs}
%

%
%

%





\end{document}